	\documentclass[12pt]{article}%
	\usepackage{epsfig,psfrag,subfigure}
	\usepackage{amsmath,amssymb,amsfonts,latexsym, mathrsfs}
	\usepackage{graphicx,graphics}
	\usepackage{wrapfig}
	\usepackage{geometry}    % See geometry.pdf to learn the layout options. There are lots.
	\usepackage{colortbl,slashbox,booktabs,multirow}
	\usepackage[table]{xcolor}
	\usepackage{epstopdf}

   \usepackage[colorlinks=true]{hyperref}
	\providecommand{\U}[1]{\protect\rule{.1in}{.1in}}
	\providecommand{\U}[1]{\protect\rule{.1in}{.1in}}
	\providecommand{\U}[1]{\protect\rule{.1in}{.1in}}
	\providecommand{\U}[1]{\protect\rule{.1in}{.1in}}

	\providecommand{\U}[1]{\protect\rule{.1in}{.1in}}

\begin{document}

	\input{defn.hyb}
	\input{defn.eqs}
	\def\bEs{\mbox{\boldmath $\mathcal{E}$}}
\def\tpsi{\widetilde{\psi}}
   \def\olC{\mbox{$\overline{C}$}}
   \def\oG{\mbox{$\overline{G}$}}
	\def\bj{\mbox{\boldmath$j$}}
   \def\omu{\mbox{$\overline{\mu}$}}
   \def\orho{\mbox{$\overline{\rho}$}}
   \def\olpi{\mbox{$\overline{\pi}$}}

	\title{Macroscopic Flow Potentials in Swelling Porous Media}
	\author{Lynn Schreyer-Bennethum
	}
	\maketitle

	\noindent
{\bf ABSTRACT}

\vspace{\baselineskip}

\footnotetext{Lynn Schreyer-Bennethum, Campus Box 170, 1250 14th Street Sixth Floor, Denver, CO 80202}

In swelling porous media, the potential for flow is much more than
pressure, and derivations for flow equations have yielded a variety
of equations.    In this paper we show that the macroscopic flow
potentials are the electro-chemical potentials of the components
of the fluid and that other forms of flow equations, such
as those derived through mixture theory or homogenization, are a result
of particular forms of the chemical potentials of the species.
It is also shown that depending upon whether one is considering
the pressure of a liquid in a reservoir in electro-chemical equilibrium
with the swelling porous media, or the pressure of the vicinal liquid
within the swelling porous media, a critical pressure gradient threshold
exists or does not.

\vspace{\baselineskip}


%\layout

%\vspace{24pt}
\vspace{2\baselineskip}

\noindent
\section{Introduction}

\vspace{\baselineskip}

Swelling porous materials are ubiquitous - they
occur in soils such as swelling clays (montmorillonite),
biotissues (cartilage), and in drug delivery
systems such as Aleve (swelling polymers).   Experiments
are performed at the {\em microscale} (scale at which the solid
and liquid or adsorbed liquid can be distinguished)
and at the {\em macroscale} (scale at which the swelling porous
media appears to be homogeneous, i.e.\ one cannot
distinguish between the phases).      The concept
of {\em pressure} at each of these scales are often
confused and interchanged.  Example of terms used include 
'disjoining pressure', 'osmotic pressure', and 'swelling pressure'  and
are attributed to
the double-layer forces, van der Waals dispersion
forces, osmotic forces, and surface hydration forces.  

In addition, it is unclear which microscopic
forces are dominant to macroscopic behavior.  For example,
although direct measurements indicate surface hydration is considered 
to have short-range effects (up to 4 monolayers of water entering
between layers of montmorillonite clay) 
\cite{vanOlphen,bergeron}, experiments by Low \cite{low4,low3}
indicate that
the macroscopic affects of these interactions
can explain the osmotic swelling of montmorillonite soils
in which the swelling is due to 100's of layers of
water.

Several upscaling approaches have been used to
arrive at a macroscopic model for flow through a swelling
porous media, and with these approaches a 
variety of definitions of macroscopic flow potentials.  
It is the purpose of
this paper to propose a macroscopic form for flow, derived
from a hybrid mixture theory formulation \cite{becu02,becu02b}, and 
demonstrate how the form involving electrochemical potentials is
a generalization of equations derived using
homogenization \cite{murad06a}, and is consistent with a 
Lagrangian mixture theoretic approach \cite{huyghe97,huyghe97b}.  
In the process
we illustrate that a pressure gradient threshold may exist,
depending upon how the pressure is measured.

For simplicity we assume that the swelling porous medium
is composed of a solid and liquid phase (i.e.\ no gaseous phase).
The solid phase (polymer, montmorillonite) is assumed to be negatively
charged and the fluid contains cations, ions, and a neutral
liquid.  We will refer to the liquid phase
as vicinal fluid to distinguish it from the bulk phase (liquid
unaffected by its vicinity to the solid phase, or reservoir fluid).

In the first section we review the microscale forces.  We next review
macroscopic quantities: osmotic repulsion, surface
hydration, and disjoining pressure.  We derive the flow
equation in terms of chemical potentials from hybrid mixture theory
results and discuss pressure gradient thresholds.   Then we illustrate
how the potential form of the flow equation can be used to derive
forms derived via homogenization (Moyne and Murad
\cite{murad06a}) used to model swelling montmorillonite, and is consistent 
with the mixture theory approach of Huyghe and Janssen
\cite{huyghe97} used to model swelling biotissues \cite{huyghe97b}.    
Although the models appear
quite different, we show they can be derived from the potential
form under particular assumptions on the chemical potentials of
the species.

\section{Microscale Forces}

At the microscale there are various forces, some
attractive and some repulsive, that cause a swelling porous
medium to swell  (repulsive forces dominate) or shrink
(attractive forces dominate).  
In this section we summarize
some of the forces considered to be dominant 
for determining the behavior at the macroscale.   
We note that these different categories of forces
are ambiguous and not disjoint.

{\bf Electrostatic repulsion:}  Due to the solid
phase being e.g.\ negatively charged, the cation and anion
fields at the microscopic scale in the vicinal
fluid are neither equal nor uniform, and as a result, there is 
a microscopically varying electric field.  One could
solve for the electrostatic condition coupled
with diffusion of ions (Poisson-Boltzmann equation \cite{newman,vanOlphen}),
but for practical situations, the complexity of the 
microstructure makes this a difficult task.  The repulsion
forces become significant when the increased cation/anion
concentrations near each surface (the double layer consisting of first the
cation dominant layer and then the anion dominant layer) begin 
to interact (double-layer
overlap) due to the proximity of the two surfaces.   These are 
considered to dominate at long-range scales.

{\bf Van der Waals attraction:}  This is an attractive force
acting between {\em all} atoms and molecules, regardless
of whether they are charged or uncharged \cite{israel}.  
%A Dutchman named van der Waals modified the Ideal gas
%law, $PV = nRT$ to apply for more general gases:
%$(P+a/V^2)(V-b)=nRT$ where $b$ accounts for the finite size
%of molecules and $a/V^2$ accounts for intermolecular 
%attractive forces which act as a virtual addition to the pressure.
The current trend is to label any additional non-pressure forces not attributed
to electrostatic forces as
Van der Waals: London forces, dispersion forces, charge-fluctuation
forces, and induced-dipole induced-dipole forces \cite{israel}.
Although some of these listed forces may be repulsive forces,
the net Van der Waals forces are considered to be attractive
and act on a shorter spatial scale than electorstatic but
not as short as surface hydration forces \cite{vanOlphen}.

DLVO theory, named after its founders, Derjaguin, Landau, Verwey,
and Overbeek \cite{derj1,verwey} incorporates electrostatic double-layer forces
and van der Waals dispersion and was developed to describe particle interactions 
\cite{bergeron}. This model has been criticized \cite{mcbride}.
Although both forces contributing to DLVO (electrostatic and
van der Waals) are static in nature, these two forces often 
equilibriate rapidly relative to other forces and so this assumption
is appropriate for many systems
\cite{israel}.

\section{Macroscale Forces}
Many experiments are performed at the {\em macroscale}, i.e.\ scale
at which one cannot distinguish between the liquid and solid phases.
Terms used at this scale include osmotic repulsion, surface hydration,
and disjoining pressure.

{\bf Osmotic Repulsion:}  
Osmotic repulsion is the force that measures how different
species interact, and is usually measured through the
osmotic pressure experiment. For example, consider a solution
(e.g.\ water and sugar) separated by a semipermeable membrane that
allows water but not sugar to pass through. There is a difference
in height and this is related to the osmotic pressure (technically
one has to take into account the effect of the membrane but for
the purposes of this paper we will consider it to be the difference
in height).   The {\em osmotic pressure}, $\pi$, is the pressure
that must be applied to the mixture to stop
the influx of solvent \cite{atkins,castellan}.    This definition holds whether
one species is charged or not.

We can derive an expression for the osmotic pressure.  In an
osmotic pressure experiment, the chemical potential on
either side of the membrane is equal at equilibrium.
For a component of a liquid solution which behaves as an
ideal gas in the gaseous phase, the chemical potential is given by
\cite{atkins} (see also Appendix B):
\begin{eqnarray} \label{e1.10}
    \mu^\lj(T,p,C^\lj) = 
    \mu^\lj_p(T,p) + \frac{RT}{m^j} \ln \left(\frac{p^{g_j}}{p^{g_j}_m}\right),
\end{eqnarray}
where $\mu^\lj$ is the mass chemical potential (energy per mass) of
species $j$ in the liquid phase, 
$C^\lj$ is the mass concentration of species $j$ in 
the mixture,
$\mu^\lj_p$ is the mass chemical
potential of pure species $j$ at the same temperature and
pressure in the liquid phase, $R$ is the universal gas constant,
$T$ is the absolute temperature, $m^j$ is the molar mass (mass of
one mole of $j$), $p^{g_j}$ is the partial pressure of species $j$
in the gaseous phase in equilibrium with the mixture, and
$p^{g_j}_m$ is the maximum partial pressure of species $j$ in the
gaseous phase obtained when in equilibrium with pure species $j$ in the liquid
phase.  The {\em activity}, $a^\lj$ is defined as the ratio of these
two partial pressures, $a^\lj = p^{g_j}/p^{g_j}_m$.
If the liquid mixture is ideal (so that {\em Raoult's law} applies)
then the activity may be replaced with the molar
concentration, $x^j$.  

Let's assume that on one side of the membrane the mixture is pure
solvent (e.g.\ water), which we label the $N$th component,
and does not contain species $j$ and on the 
other side the mixture contains species $j$ and solvent.  On the
side of the mixture the pressure will be higher, by an amount
proportional to the osmotic pressure, $\pi^\lj$.  
The chemical potential of the solvent must be equal on both
sides and we have
\begin{eqnarray} \label{e1.12}
    \mu^{l_N}_p (T,p) = \mu^{l_N}(T,p+\pi^\lj,C^\lj)
    = \mu^{l_N}_p(T,p+\pi^\lj) + \frac{RT}{m^N} 
    \ln(a^\lj) .
\end{eqnarray}
To evaluate $\mu^{l_N}_p(T,p+\pi)$, we begin with the total differential
\begin{eqnarray} \label{e1.13a}
    d\mu^{l_N}_p = \frac{\p \mu^{l_N}_p}{\p T} dT + 
    \frac{\p \mu^{l_N}_p}{\p p} dp.
\end{eqnarray}
For a pure substance, $\frac{\p \mu^\lj_p}{\p p} = 1/\orho^\lj$
where $\orho^\lj$ is the specific density of species $j$ in
the liquid phase with units of mass of $j$ per volume of $j$ (see
Appendix A, or \cite{atkins,castellan}).
Integrating at constant temperature from the state at pressure $p$ to
the state where pressure is $p+\pi$ we get
\begin{eqnarray} \label{e1.13b}
    \mu^{l_N}_p(T,p+\pi^\lj) - \mu^{l_N}_p(T,p) = \int_{p}^{p+\pi^\lj} 
    \frac1{\orho^{l_N}} dP.
\end{eqnarray}
Using this expression in (\ref{e1.12}b) to eliminate 
$\mu^{l_N}_p(T,p+\pi^\lj)$ and then subtracting
$\mu^{l_N}_p(T,p)$ from both sides
gives:
\begin{eqnarray} \label{e1.14}
     \int_p^{p+\pi^\lj} \frac1{\orho^{l_N}} dP = -\frac{RT}{m^N} 
     \ln(a^\lj).
\end{eqnarray}
If the density of the solvent, $\rho^{l_N}$, is constant, then we have
\begin{eqnarray} \label{e1.15}
     \frac1{\orho^{l_N}}\pi^\lj = -\frac{RT}{m^N} 
     \ln(a^\lj),
\end{eqnarray}
and further if we have an ideal solution \cite{atkins,castellan}, then
\begin{eqnarray} \label{e1.16}
    \pi^\lj = -\frac{RT\orho^{l_N}}{m^N} 
    \ln(x^{l_N}),
\end{eqnarray}
where $x^{l_N}$ is the molar fraction of solvent, given by moles of solvent
per moles of mixture.   If the soution is dilute, so that $x^{l_N} =
1-x^\lj$ where $x^\lj$ is small, then $\ln(x^{l_N}) \approx -x^\lj$ and
approximating the number of moles of $N$ as being equal to the moles
in the solution, we get
\begin{eqnarray} \label{e1.17}
    \pi^\lj \approx RT C_m^\lj
\end{eqnarray}
where $C_m^\lj$ is the molar concentration of $j$ (moles of $j$ per
moles of solution), which is known as the {\em van't Hoff equation}.

{\bf Surface Hydration:}  These are short-range bonding forces between the 
solid surface and the water that causes
one to ten layers of water to be held tightly \cite{israel,vanOlphen}.
These forces perturb the vicinal liquid, so that
it behaves differently from its bulk-phase counterpart 
(water free of adsorptive forces) 
\cite{low4,grim}.  Experiments by Low 
\cite{low4} 
indicate that the macroscopic effects of these interactions can qualitatively
completely account for many macroscopic experimental results.   
%In particular, changing the pH (CHECK) of the liquid, does not
%affect the shape of the curve, only the shift (CHECK).

\begin{figure}[bt]
\begin{center}
\includegraphics[height=1.5in]{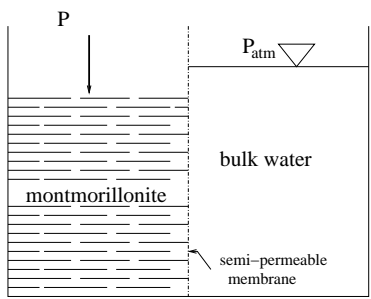}
\caption{Reverse osmotic swelling pressure experiment.}
\label{fig:low2}
\end{center}
\end{figure}

If one considers the solid-liquid mixture as a mixture itself, then
the hydration forces can be thought of as the osmotic force of the
solid particles.  In fact the reverse osmotic swelling pressure
experiment is presicely what was done by Low \cite{low4} for
montmorillonite soils.
In this experiment (see Figure \ref{fig:low2}), the liquid mixture and liquid mixture with well-layered
clay minerals were separated by a semipermeable membrane which did not
allow the clay minerals to penetrate, and the pressure
required to keep the clay mixture from swelling was measured.  In
this case, the quantity of clay mineral was measured in terms of
the distance separating the clay platelets ($\lambda^l$) which represents
the concentration of the clay mineral.  In this case, the hydration
pressure was determined experimentally to be exponentially related to the clay mineral 
concentration.  If $\lambda^s$ is the thickness of the clay plates
then \cite{low4}
\begin{eqnarray} \label{e1.18}
    \pi = p_0 e^{\frac{\lambda^s}{\lambda^l}} - p_0,
\end{eqnarray}
where $p_0$ is the reference (atmospheric) pressure.
Equation (\ref{e1.18}) was also obtained via hybrid 
mixture theory, \cite{ach}.
Note that this result is quite different from a pure liquid mixture in 
which the
osmotic pressure is proportional to the log of the concentration,
(\ref{e1.16}), but they have the same general shape - as the
moisture goes to zero, the swelling pressure goes to infinity, and
as the moisture content goes to 1 ($\lambda^s = 0$), the swelling pressure
goes to zero.

%\begin{wrapfigure}[15]{r}{0.6\textwidth}
%\end{wrapfigure}

\begin{wrapfigure}[15]{r}{0.6\textwidth}
\begin{center}
\includegraphics[width=8cm]{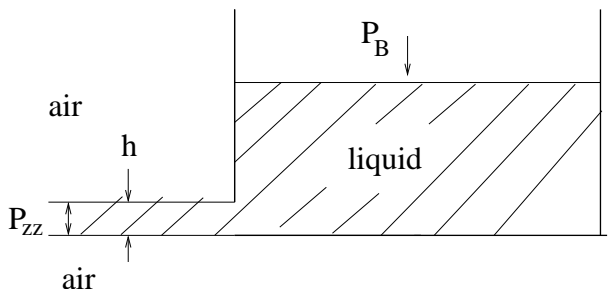}
\caption{Cartoon illustrating variables used to determine
disjoining pressure.}
\label{fig:disjoin}
\end{center}
\end{wrapfigure}

{\bf Disjoining pressure:} is a concept traditionally used in the field
of foams (gas-liquid dispersions) and emulsions (liquid-liquid dispersions),
where the stability of the system relies on the stability 
of the thin liquid films \cite{bergeron}.  
If two interfaces
(in the case of foams, air-liquid and liquid-air) are separated
by a distance $h$, then if $h$ is small enough there
is no portion of the interlayer (i.e.\ liquid film) which
possesses the properties of the bulk fluid (see Figure \ref{fig:disjoin}).  
In such a case,
Derjaguin and Churaev \cite{derj3} state
\begin{quote}
in mechanical
equilibrium the disjoining pressure, $\pi(h)$, is equal to the difference
existing between the component, $P_{zz}$ of the
pressure tensor in the interlayer and pressure, $P_B$, set up in
the bulk of the phase from which it has been formed by thinning
out:
\begin{eqnarray} \label{e2.10}
\pi(h) = P_{zz} - P_B = P_N-P_B.
\end{eqnarray}
In the simplest case of a one-component liquid phase, mechanical
equilibrium under isothermic conditions implies thermodynamic
equilibrium.  In that case the disjoining
pressure is a single-valued function of the interlayer thickness,
$h$,..."  
\end{quote}
This definition has been extended so that it
applies to curved surfaces by Kralchevsky and Ivanov \cite{kiv}.
This mechanical definition is thought to be equivalent to the
thermodynamic definition in terms of the Gibbs free energy, $G$, as
\cite{bergeron,eriksson}
\begin{eqnarray} \label{e2.11}
\pi(h) = - \left.\frac{\p G}{\p h}\right|_{T,P,A,N_i}
\end{eqnarray}
where the variables held fixed while taking the
partial derivative include temperature, $T$,
pressure, $P$, the area of the interface, $A$, and
the number of moles of each constituent making up
the thin film, $N_i$.

According to Bereron \cite{bergeron}, 
the disjoining pressure
is thought to be due to many forces:  electrostatic double-layer,
van der Waals dispersion forces, short-range structural forces
such as hydration, and other forces.   In the field of 
thin liquid soaps, most treat these forces as being additive,
although it is not clear that this is a valid assumption
\cite{attard88,attard88b}. 
More than one author has come to the conclusion that
the swelling pressure and average disjoining pressure
are the same, e.g.\ \cite{derj1} (p.\ 282).

\section{Flow in Terms of Chemical Potentials}

We begin with a formulation developed using hybrid mixture theory
(HMT) \cite{benn2,benn3}.   In this approach,
the microscale field equations (conservation
of mass, linear and angular momenta, energy, and electroquasistatic
form of Maxwell's equations) are volume averaged
to produce macroscopic quantities and equations, and then macroscopic
constitutive equations are obtained by assuming a set of constitutive
variables are a function of the same set of (macroscopic) independent variables
and then exploiting the entropy inequality  in the spirit of
Coleman and Noll \cite{coleman}.  
This approach has the advantage of developing constitutive equations
directly at the macroscale, however coefficients in the macroscopic
constitutive equations are not directly linked to microscopic quanitities.
In what is presented here, the only geometric information
retained at the macroscale is the volume fraction, although
this approach can be expanded to incorporate e.g.\ interfacial
surface density \cite{becu4,hgd}.
In \cite{becu02b} the independent variables included
\begin{eqnarray} \label{e3.11}
    \e^l,\; T,\; \rho^\aj,\; \bv^{l,s},\; \bEs^s,\; \bE,\; z^\aj,\; 
\bnab\e^l,\; \bnab T, \; \bnab \rho^\aj,...
\end{eqnarray} 
where $\e^l$ is the volume fraction of the liquid phase, $T$ is temperature,
$\a$ represents the phase ($\a=l$ for liquid and $\a=s$ for solid), 
$\rho^\aj$ is the density of the $\a$-phase (mass of species $j$ in the
$\a$ phase per unit volume of the $\a$ phase), 
$\bv^{l,s}$ is the velocity of the liquid relative to the solid phase,
$\bEs^s$ is the strain of the 
macroscopic (smeared out) solid phase, $\bE$ is the electric 
field, and $z^\aj$ is the fixed
charge density associated with species $j$ of phase $\a$.

The thermodynamic definition of liquid pressure is given by
\begin{eqnarray}\label{e3.12}
p^l = \sum_{j} \rho^l\rho^\lj \left.\frac{\p \psi^l}{\p \rho^\lj}\right|_{\e^l,...}
\end{eqnarray}
where $\psi^l$ is the intensive (per unit mass) Helmholtz potential.
One can either enforce electroneutrality with a Lagrange multiplier,
$\Lam$, or include an electric field.  In the former approach,
$\Lam$ is the streaming potential.  In \cite{benn3} it is shown 
that $p^l+q_e^l \Lam$ where 
$q_e^l$ is the charge density and $\Lam$ is a streaming potential,
is related to one third the trace of the macroscopic liquid cauchy
stress tensor - thus the thermodynamic definition is related
to what is physically measured \cite{bewe03}.

Another pressure, 
the "swelling pressure", is thermodynamically
defined as:  
\begin{eqnarray} \label{e3.14}
\pi^l = \e^l \rho^l \left.\frac{\p \psi^l}{\p \e^l}\right|_{\rho^l,...},
\end{eqnarray} 
where $\e^l$ is the liquid volume fraction and where the partial
derivative is evaluated
keeping the other independent variables (density, concentrations, temperature)
fixed.   It is defined so that it is a positive quantity for
a swelling mixture.  Clearly this is a macroscopic form of the thermodynamic
definition of the disjoining pressure, and in fact, if the solid
phase is structured so that it does not support stress (e.g.\ parallel 
platelets), it can be shown \cite{bewe03} that for
a single component liquid,
\begin{eqnarray} \label{e3.16}
\pi^l = -\e^l \left.\frac{\p p^l}{\p \e^l}\right|_{G^l,...}, 
\end{eqnarray}
where the partial 
derivative
is evaluated keeping the Gibbs potential (chemical potential) fixed,
which is exactly the reverse-osmotic swelling potential experiment
used to measure the osmotic force.   Note that if the material
is not swelling, then the energy of the liquid phase would not
change with liquid content and the swelling pressure is zero.

It can be shown
that $p^l$ and $\pi^l$ are related via a third thermodyanic
property which is related to the change in Helmholtz potential with respect
to volume keeping the mass fixed \cite{bewe03}:
\begin{eqnarray}\label{e3.18}
p^l = -\e^l\rho^l\left.\frac{\p\psi^l}{\p\e^l}\right|_{\e^l\rho^l,...} + \pi^l
\end{eqnarray}
Equation (\ref{e3.18}) is mathematically exact (no assumptions), and if one
converts to extensive variables one can show that this new quantity is the 
traditional thermodynamic definition of pressure:  change in energy with 
respect to volume keeping the mass fixed.  Thus the pressure in the liquid
phase has two components:  one which is the 'classical' pressure for a bulk
fluid, and the other the swelling pressure \cite{bewe03}.  If the swelling
pressure is zero, then the traditional thermodynamic pressure is the
same as one third the trace of the cauchy stress tensor of a liquid.

Assuming: (i) terms involving the polarization vector
field are negligible, (ii) the gravitational term is negligible, 
(iii) isothermal conditions, (iv) sufficient moisture so that
the liquid phase does not support shearing forces, (v) the charge
associated with each species, $z^j$, is fixed, and (v)
{\em not} assuming charge neutrality, 
the resulting Darcy's law using Hybrid Mixture Theory is given by
\cite{becu02b}
\begin{eqnarray} \label{e3.20}
    \bR \cdot \bv^{l,s} &=& -\e^l \bnab p^l - \pi^l\bnab \e^l
 + \e^l q_e^l \bE 
- \sum_{j=1}^N r^\lj \bv^{\lj,l} 
\\ \label{e3.21}
&=& -\e^l \rho^l \bnab G^l + \sum_{j=1}^N \e^l (\rho^l)^2 
   \frac{\p \psi^l}{\p \rho^\lj} \bnab C^\lj
   + \e^l q_e^l \bE
- \sum_{j=1}^N r^\lj \bv^{\lj,l} 
\end{eqnarray} 
where $q^l_e$ is the charge density of the liquid phase,
$G^l = \psi^l - p^l/\rho^l$ is the Gibbs potential for the liquid phase,
$C^\lj$ is the mass concentration (mass of species $j$ in the liquid
phase per mass of liquid phase), and $\bv^{\lj,l} = \bv^\lj-\bv^l$ is
the diffusive velocity.   The last term involving the diffusive velocities
captures the effects of ion hydration and relative friction
between the mass-averaged velocity and species velocity.  If the
diffusive velocities ($\bv^{\lj,l}$) are small then this term
may be neglected.
Note that in (\ref{e3.20}) there
are no terms directly involving chemical potential or concentrations of species
that contribute to flow.  

Changing the concentrations of the species
making up the liquid phase changes the pressure through (\ref{e3.12})
and through relative velocities.   The generalized version of Fick's law
for diffusion is given by \cite{bmc2,newman}
\begin{eqnarray} \label{e3.22}
    \bv^{\lj,l} = \bQ^\lj \cdot \bnab \mu^\aj
\end{eqnarray}
%Newmann Section 12.6
where $\bQ^j$ is a diffusion coefficient tensor which may be a function of
volume fraction, temperature, and densities so that
\begin{eqnarray} \label{e3.23}
    &\bR \cdot \bv^{l,s} &= -\e^l \bnab p^l - \pi^l\bnab \e^l
 + \e^l q_e^l \bE 
- \sum_{j=1}^N r^\lj \bQ^\lj  \cdot \bnab \mu^\lj 
\\ \label{e3.23b}
&&= -\e^l \rho^l \bnab G^l + \sum_{j=1}^N \e^l (\rho^l)^2 
   \frac{\p \psi^l}{\p \rho^\lj} \bnab C^\lj
   + \e^l q_e^l \bE
   \nonumber \\ && \hspace{0.5in}
- \sum_{j=1}^N r^\lj \bQ^\lj\cdot \bnab \mu^{\lj,l} 
\end{eqnarray}

From equation (\ref{e3.23}) we see that if $\pi^l$ is not zero
we have a pressure gradient threshold - i.e.\ gradient in the
volume fraction can offset the pressure in the liquid pressure until
$\pi^l \bnab \e^l$ is maximum, and then further increasing the
pressure gradient will induce flow \cite{sanchez07,wei09,song10}.   
An analagy between this
and concentration gradients can be made - flow is induced by
a ``concentration'' gradient of the solid phase.

Next we express the flow equation in terms of liquid chemical potentials
because the electro-chemical potentials are continuous between vicinal 
and bulk fluids and because it may
be more useful for numerical solutions
\cite{murad06a,huyghe,huyghe97}.

Within HMT, the chemical
potential is given by \cite{callen2,bmc2}:
\begin{eqnarray} \label{e3.15}
\mu^\aj = \frac{\p(\rho^\a\psi^\a)}{\p \rho^\aj}
\end{eqnarray}
whereas the electrochemical potential \cite{newman} is given by
\begin{eqnarray} \label{e3.15b}
\tmu^\aj = \mu^\aj + z^\aj \phi
\end{eqnarray}
where $\phi$ is the electric field potential and $z^\aj$ is
the charge density (per unit mass) for species $j$ in phase $\a$.

We consider two cases:
one in which the liquid (and bulk) phase is composed of only
one constituent, and then a multi-constituent liquid phase.

First consider a liquid phase which is composed of only one constituent.
The relationship between the Gibbs potential and chemical
potentials is given by \cite{callen2}:
\begin{eqnarray} \label{e3.25}
G^\a = \sum_{j=1}^N C^{\alpha_j} \mu^{\alpha_j} \qquad \a = l,B.
\end{eqnarray} 
So for a single component phase, the concentration is 1 
and $\bnab C^\lj = \bo$. Thus all diffusion velocities are zero, and the 
Gibbs potentials for the vicinal and bulk phases are equal up to 
the Lorentz term:  
$G^l+q_e^\a \phi =G^B$ where we assume the bulk phase fluid is charge neutral
($\bnab\phi^B = \bo$).   Using the relationship between the Gibbs potential
and the Helmholtz potential, $G=\psi - p/\rho$ we get that the 
right-hand side of Darcy's equation (\ref{e3.23b}), not including
the hydration terms, is given by:
\begin{eqnarray} \label{e3.26}
&& -\e^l\rho^l \bnab G^B +\e^l q_e^l\bE  \\
&& \hspace{0.5in} = -\e^l \rho^l \bnab \psi^B + \e^l \rho^l \frac{p^B}{(\rho^B)^2}
\bnab \rho^B - \e^l\rho^l\frac1{\rho^B} \bnab p^B 
   + \e^l q_e^l \bE.
\end{eqnarray} 
Now assume the bulk phase Helmholtz potential 
is only a function of density.  Then using
the thermodynamic definition of pressure, (\ref{e3.12}), the
flow equation can be written as
\begin{eqnarray} \label{e3.27}
    \bR \cdot \bv^{l,s}&=& -\e^l\rho^l \frac{\p \psi^B}{\p \rho^B}\bnab \rho^B
 + \e^l \rho^l \frac{p^B}{(\rho^B)^2}
\bnab \rho^B - \e^l\rho^l\frac1{\rho^B} \bnab p^B
   + \e^l q_e^l \bE.
   \\ \label{e3.27b}
  &=& -\frac{\e^l\rho^l}{\rho^B} \bnab p^B 
   + \e^l q_e^l \bE.
\end{eqnarray} 
and we see that if we write the flow equation in terms of potentials 
of the vicinal fluid as in equation (\ref{e3.20}), we have both a pressure
and volume fraction potential, but if the flow
equation is written as the bulk phase we have only
a pressure potential.   {\em This implies that if one is measuring
a vicinal pressure, there may be a pressure gradient threshold, but
if one is measuring pressure of the bulk phase, there is no
pressure gradient threshold.} 

Now consider a multi-component liquid phase.  
Beginning with the right-hand side of Darcy equation (\ref{e3.23b})
and using (\ref{e3.25}) we have:
\begin{eqnarray} \label{e3.30}
-\e^l\rho^l \sum_{j=1}^N \bnab (C^{l_j} \mu^{l_j}) 
+ \sum_{j=1}^N \e^l(\rho^l)^2 \frac{\p \psi^l}{\p \rho^\lj} \bnab C^\lj
   + \e^l q_e^l \bE
- \sum_{j=1}^N  r^\lj \bQ^\lj \cdot \bnab \mu^\lj.
\end{eqnarray}
Using (\ref{e3.15}) for the thermodynamic definition of the chemical
potential to eliminate $ \frac{\p \psi^l}{\p \rho^\lj} $ we have (\ref{e3.30})
is equivalent to
\begin{eqnarray} \label{e3.31}
&& -\e^l\rho^l \sum_{j=1}^N C^\lj \bnab \mu^{l_j} 
-\e^l\rho^l \sum_{j=1}^N \mu^\lj \bnab C^{l_j} 
+ \sum_{j=1}^N \e^l\rho^l \left(  \mu^\lj - \psi^l \right) \bnab C^\lj
\\ && \hspace{1.0in}
   + \e^l q_e^l \bE
- \sum_{j=1}^N r^\lj \bQ^\lj \cdot \bnab \mu^\lj.
\end{eqnarray} 
Noting that $\sum_{j=1}^N C^\lj = 1$ so that the second part of the
third term on the right side is zero and that the electrochemical potentials
between the vicinal phase and bulk phases are equal, 
we can rewrite the flow equation as:
\begin{eqnarray} \label{e3.33}
    \bR \cdot \bv^{l,s} &=&
    % \nonumber \\ && \hspace{0.2in}
    -\e^l\rho^l \sum_{j=1}^N C^\lj \bnab \mu^{l_j} 
   + \e^l q_e^l \bE
   - \sum_{j=1}^N r^\lj \bQ^\lj \cdot \bnab \mu^{B_j}
\nonumber \\  
&=& - \sum_{j=1}^N \e^l \rho^\lj \bnab \mu^{B_j}
   + q_e^B \bE
   - \sum_{j=1}^N  r^\lj \bQ^\lj \cdot \bnab \mu^{B_j}.
\end{eqnarray} 
So in a multi-component fluid, it is {\em the electrochemical potentials of
the liquid phase species}, $\tmu^\aj = \mu^\aj +z^\aj \phi$,
that are the primary potential for fluid flow.

To get one final form of this equation, consider the form of
chemical potential for a liquid to be (see Appendix B):
\begin{eqnarray} \label{e3.40}
    \mu^\lj = \mu^\lj_p(T,p_0) + \frac1{\rho^\lj_0}(p^l-p_0)
      + \frac{RT}{m^j}\ln{a^j},
\end{eqnarray}
where $\rho^\lj_0$ is the specific density of species $j$ in the
liquid phase (mass of $\lj$ per volume of $\lj$) and it is assumed 
that each component of the liquid phase is 
incompressible.
Then flow equation (\ref{e3.33}) in terms of bulk variables can
be written as:
\begin{eqnarray} \label{e3.45}
     \bR \cdot \bv^{l,s} &=&
 -\bnab p^B - \sum_{j=1}^N \e^l \rho^\lj 
    \frac{RT}{m^j a^j}\bnab a^j
   + q_e^B \bE
   - \sum_{j=1}^N r^\lj \bQ^\lj \cdot \bnab \mu^{B_j}.
\end{eqnarray} 
where we used the fact that $\displaystyle \sum_{j=1}^N 
\frac{\e^l\rho^\lj}{\rho^\lj_0}=1$.  So in the reservoir bulk fluid
the primary driving forces are the activities (which are closely related
to concentrations) and just as we had
for a single-component fluid, the bulk phase pressure. Hydration
of ions is an additional component whose magnitude for
many problems has yet to be determined.

\begin{figure}[b]   % b= bottom of page
    \begin{minipage}[b]{0.3\linewidth}  % b= bottom of minipage
        \epsfysize=1.2in  % height of figure
        \centerline{
        \epsfbox{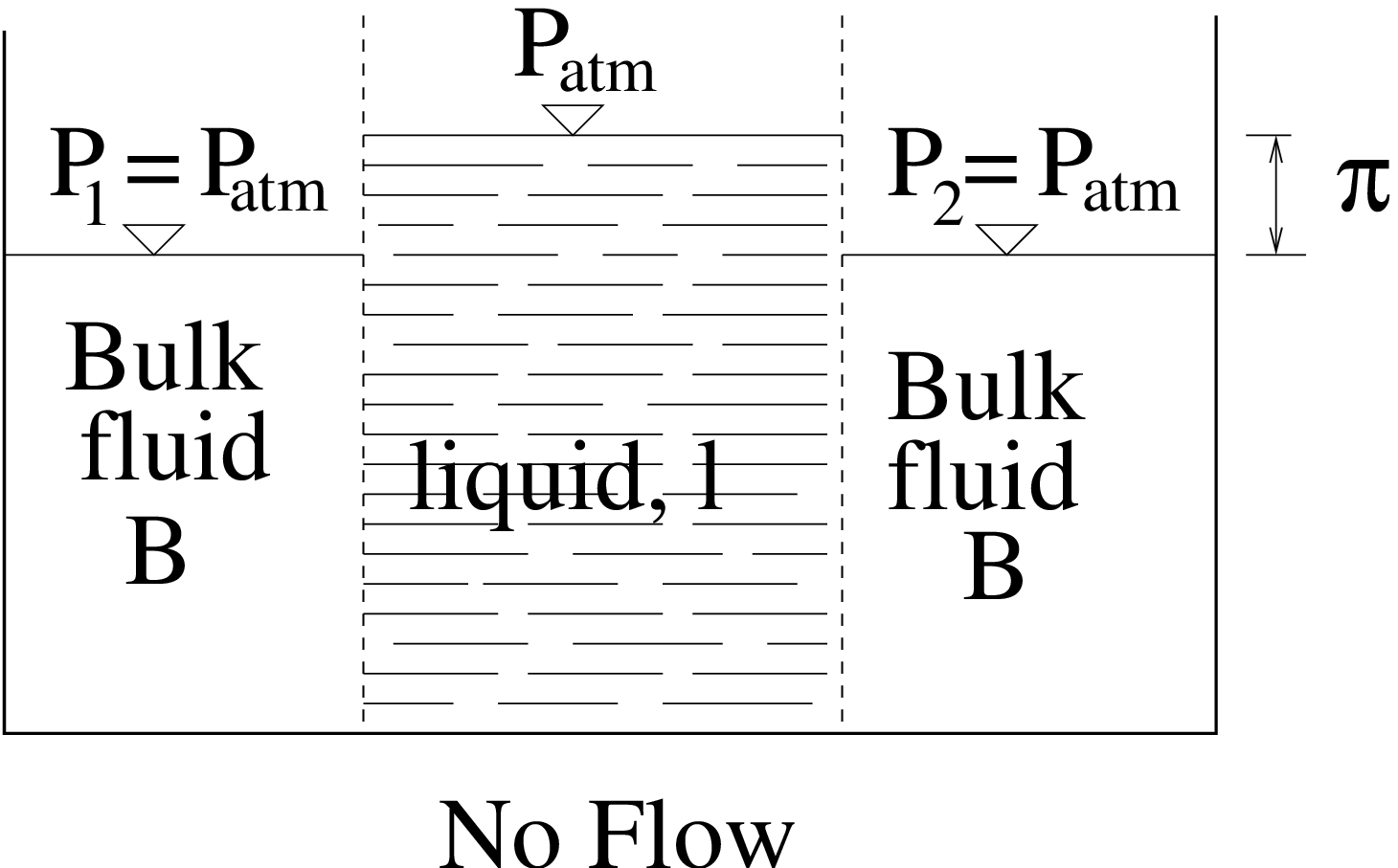}}
        \caption{equal chemical potentials}
        \label{fig:f5a}
    \end{minipage}
    \hspace{0.2in}
    \begin{minipage}[b]{0.3\linewidth}  % b= bottom of minipage
        \epsfysize=1.2in  % height of figure
        \centerline{
        \epsfbox{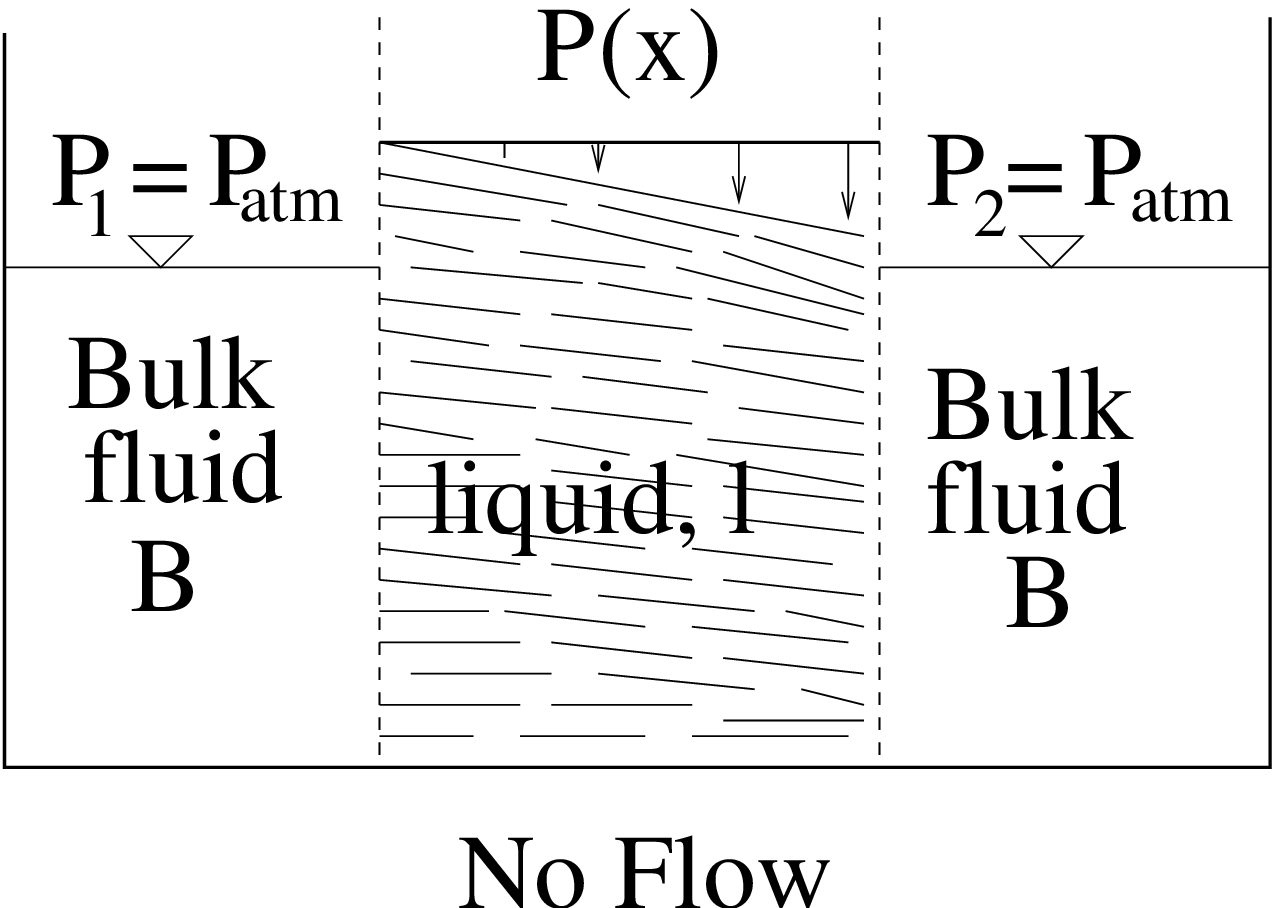}}
        \caption{$p^l$ is a function of $x$}
        \label{fig:f5b}
    \end{minipage}
    \hspace{0.2in}
    \begin{minipage}[b]{0.3\linewidth}  % b= bottom of minipage
        \epsfysize=1.2in  % height of figure
        \centerline{
        \epsfbox{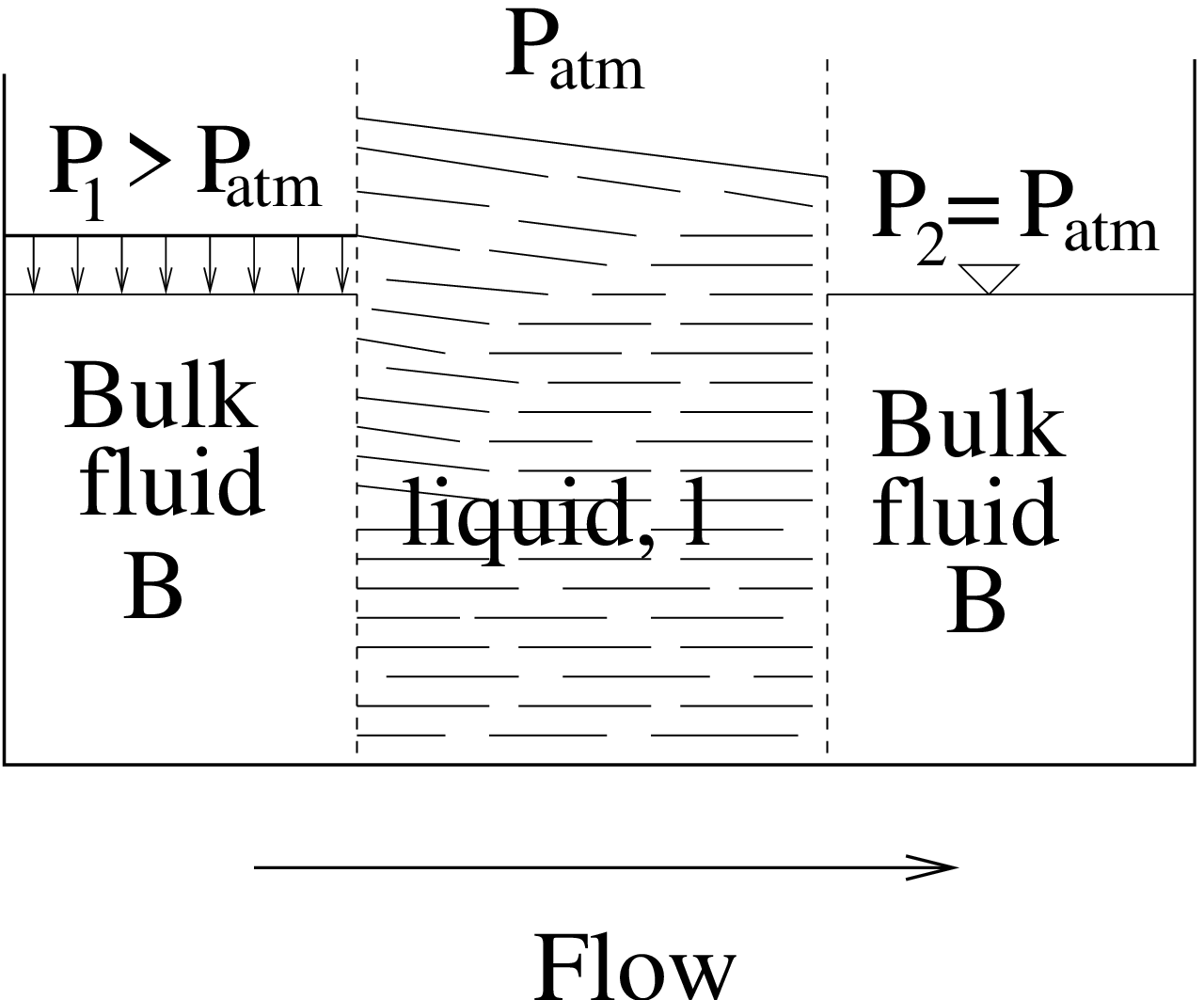}}
        \caption{higher chemical potential on side 1}
        \label{fig:f5c}
    \end{minipage}
\end{figure}

Next we illustrate the insight that can be obtained by writing 
the equations in terms of
the chemical potential. Consider Figures \ref{fig:f5a}, \ref{fig:f5b},
and \ref{fig:f5c},  
where we have a reverse osmotic
swelling potential experimental set up where the two bulk phases
are separated from a swelling porous material such as 
montmorillonite clay mixture by a semipermeable
membrane.  Across the membrane the electro-chemical potentials are
continuous \cite{callen2,newman}.    
Assume the chemical potential
of the bulk phase is determined solely by the bulk phase pressure
and the chemical potential of the liquid in the clay mixture is determined
by the pressure and the liquid volume fraction (i.e.\ gradients in
liquid concentrations and macroscale electric fields are negligible).  
In Figure \ref{fig:f5a} there is no gradient in the chemical potential of
the liquid phase, so there is no flow.  The swelling pressure, $\pi$, 
is proportional
to the difference in height of the mixture and the bulk fluid.
In Figure \ref{fig:f5b} a pressure is applied to the clay mixture.  If the
clay mixture is well-layered then (up to the hydrostatic pressure) the
applied pressure is equal to liquid pressure, $p^l$.  However in
this case the gradient in the applied pressure is offset by a gradient
in volume fraction, and because the chemical potentials in the bulk phase
(which is in chemical equilibrium with the vicinal phase) are the same,
there is no net flow.  In \ref{fig:f5c} a pressure is applied to 
the left side, changing its chemical potential.  This applied pressure
causes the clay to swell more on the left side and because there is
an overall chemical potential gradient the liquid flow is induced.
We could recreate these same pictures for a multi-component liquid
by keeping the pressure fixed and changing the concentrations.

\section{Comparison with Other Models}
In this section, we show how the chemical potential formulation
for flow is a generalization of several other models, including
that derived using homogenization (Moyne and Murad \cite{murad06a}), 
and a mixture theoretic derivation of Huyghe and Janssen \cite{huyghe97}.

\subsection{Model of Moyne and Murad} \label{s5.1}
In \cite{murad06a}, Moyne and Murad
use homogenization 
to upscale microcopic field (conservation laws and
Maxwell equations) and constitutive equations
to the macroscale.
This 
approach provides first-order equations with precise expressions
for coefficients in terms of solutions to the microscale equation
on a periodic structure. 
The microscopic equations include:  an incompressible liquid phase
(composed of a liquid, a cation and anion)
and a
linear elastic solid phase;  the conservation of momentum
with the Lorentz term added ($q^l_e\bE$ where $q^l_e$ is the charge of
the liquid (solvent) and $\bE$ is the electric field);  Gauss' law 
 assuming polarization is negligible 
($\tilde{\e}\tilde{\e}_0\bnab \cdot \bE = q^l_e$, where
$\tilde{\e}$ is the relative dielectric constant of the solvent and
$\tilde{\e}_0$ is the vacuum permittivity); the conservation
of mass for each ion, $j$,
$\p n^j / \p t + \bnab \cdot \bj^j = 0$ 
with the ion flux given by  $\bj^j=n^j\bv - Dn^j/(kT) \bnab \omu^j$
where $\omu^j$ is the chemical potential of $j$ (per molecule $j$),
$n^j$ is the volumetric concentration of ion $j$, 
$k$ is the Boltzmann's constant,
and $T$ is absolute temperature (assumed constant).  
In addition,
electroneutrality is enforced.

The solid phase is assumed to be platelet shaped (as in, e.g.\ montmorillonite).
Before upscaling, a change of variables is performed to replace
variables which may change very rapidly between the platelets (vicinal fluid)
to variables that are more smoothly varying.  With this in mind, instead
of using the chemical potentials of the ions in the vicinal liquid,
the chemical potential of the ions in the bulk fluid in 
thermodynamic equilibrium,
$\mu^j_B$, is used.  
The apparent "bulk" phase fluid is a hypothetical 
fluid in electrochemical equilibrium with the vicinal fluid \cite{coussy}.
The relationship between $\omu^j$ and $\mu^j_B$ is
obtained assuming the charged particles satisfy the Boltzman distribution, 
which itself assumes 
a single flat double layer.
Similarly, the liquid pressure is replaced
by a form of the "disjoining pressure":  
\begin{eqnarray} \label{e3.2}
\pi=P_b - p 
\end{eqnarray} 
where
$p$ is the pressure of the vicinal fluid, and
\begin{eqnarray} \label{e3.3}
P_B = \rho \mu_w = p+\int_0^\phi q(\phi) \; d\phi 
\end{eqnarray} 
is the
"local apparent bulk phase pressure", and where $\rho$ is the
density of the liquid, $\mu_w$ is the chemical potential of the
water in the liquid phase (per unit mass), and $\phi$ is the
electric potential.  
Note that $\bnab P_b = \bnab p - q_e\bE$ so
that $P_b$ incorporates the Lorentz term.
Because the relationship between $p$ and $P_B$ are assumed, 
it is not clear at this point whether $\pi$ is
the same as the mechanical definition of the disjoining
pressure as defined previously. 
After homogenizing, the resulting Darcy-type law is 
\begin{eqnarray} \label{e3.10}
\bv^0_D &=& -\bK \bnab_x p^0_b - \bK_+ \bnab_x n^{+0}_b - \bK_- \bnab_x n^{-0}_b
\\ \label{e3.10b}
 &=& -\rho \bK \bnab_x \mu^0_w - \bK_+ \bnab_x n^{+0}_b - \bK_- \bnab_x n^{-0}_b
\end{eqnarray}
where $\bK$, $\bK_-$, $\bK_+$ are second-order tensors, $\mu^0_w$ is the
chemical potential of the water in the liquid phase 
defined so that $\rho\mu_w = p_b^0$, and a superscipt
$0$ denotes the first term in a series expansion of orders $\e$. 
In this case $p^0_b$ incorporates the first-order approximation of
the Lorentz term and the remaining terms come from second-order
terms (fluctuations within the vicinal layer) and their relationship
to the Boltzman distribution.

We now show that these results are a special case of (\ref{e3.33})
Beginning with (\ref{e1.10}) and proceeding as we did to derive
(\ref{e1.13b}) we have
\begin{eqnarray} \label{e3.50}
    \mu^{B_j}(T,p^B,C^{B_j}) &=& \mu^{B_j}_p(T,p) 
    + \frac{RT}{m^j} \ln {a^{B_j}}
    \nonumber \\ 
    &=& \mu^{B_j}_0(T,p_s) + \frac1{\orho_0^{B_j}} (p^B-p_s) 
    + \frac{RT}{m^j} \ln {a^{B_j}},
\end{eqnarray}
where $p_s$ is the standard reference pressure (1 bar) and where
we assumed the specific densities, $\orho_0^{B_j}$, are constant.  At
constant temperature we thus have
\begin{eqnarray} \label{e3.51}
    \bnab \mu^{B_j} = \frac1{\orho_0^{B_j}} \bnab p^B
    + \frac{RT}{m^j a^{B_j}} \bnab a^{B_j}.
\end{eqnarray}
Let us consider that we have three species:  water, $j=w$, cations, $j=+$,
and anions, $j=-$.   If the solution is dilute then it is reasonable
that the solvent, water, will follow Raoult's law very well, so that
$a^w\approx x^{B_w} \approx 1$, so that we have from (\ref{e3.33})
\begin{eqnarray} \label{e3.53}
    \bR\cdot\bv^{l,s} = - \sum_{j=1}^N \e^l \rho^\lj \bnab \mu^{B_j} 
&=& -\sum_{j=1}^N \left[\frac{\e^l \rho^\lj}{\orho^{B_j}_0} \bnab p^B
+ \frac{\e^l\rho^\lj RT}{m^j a^{B_j}} \bnab a^{B_j} \right]
\nonumber \\  \hspace{0.5in} 
&=& - \bnab p^B -\sum_{j=+,-} \frac{n^{B_j}RT}{a^{B_j} V} \bnab a^{B_j},
\end{eqnarray} 
where $V$ is the volume of the Representative Elementary volume
and $n^B_j$ is the number of moles of $j$ in $V$.  If Raoult's law 
applies ($a^{B_j} = x^{B_j}$, then (\ref{e3.53}) has the same form 
as (\ref{e3.10}).

Comparing the two approaches through equations (\ref{e3.10b}) and 
(\ref{e3.33}), we see the results are the same up to the definition
of coefficients
if we make the following observations/assumptions: (i) in (\ref{e3.33}) 
neglect the effects
of hydration in the HMT approach, (ii) in (\ref{e3.10b}) 
recall that the definition
of $P_b$ incorporates the Lorentz term (\ref{e3.3}), (iii) in equation
(\ref{e3.33}) we can
assume the bulk chemical potential is primarily a function of the
concentrations so that $\bnab \mu^{B_j} \approx \frac{\p \mu^{B_j}}{\p C^{B_j}}
\bnab C^{B_j}$, and (iv) the coefficients $\bK$ in (\ref{e3.10b}) are a function
of the microscopic geometry and so are a function of the volume fraction.

\subsection{Model of Huyghe and Janssen} \label{s5.2}

In \cite{huyghe97}, Huyghe and Janssen
use a mixture theoretic approach in a Lagrangian framework 
to develop equations subsequently used in biological applications 
e.g.\ \cite{huyghe97b}.

They label their model the Quadriphasic model because they treat
the system as consisting of four phases: cations (+), anions (-), 
a charged solid (s), and a fluid (f).  Each ``phase'' is considered
incompressible, and that the volume fraction of the anions and cations
are negligble relative to the volume fraction of the solid and fluid phases.
Chemical interactions are neglected and electroneutrality is enforced.
A work energy function is assumed with independent variables consisting
of the Green strain, Lagrangian form of the volume fraction of the fluid
and ions, and the Lagrangian form of the relative velocities.
The generalized Darcy law derived by exploiting the entropy inequality,
neglecting inertial and gravitational terms is (equation (6) from 
\cite{huyghe97b} or equations (52) and (53) from \cite{huyghe97}):
\begin{eqnarray} \label{e3.70}
    \e^l \bv^{l,s} = -\bF\cdot\bK\cdot\bF^T \cdot\left[ \bnab(p^l-\olpi)
     + n^+\bnab\tmu^+ + n^- \bnab \tmu^- \right]
\end{eqnarray} 
where $\bF$ is the deformation tensor used to convert between eulerian
and Lagrangian frameworks, $p^l$ is the hydrodynamic pressure,
$\olpi$ is the osmotic pressure of the ions, $\tmu^\a$ is the electrochemical
potential incorporating the streaming potential, and $n^\a$ is the volumetric 
concentration of phase $\a$.   Here the osmotic pressure, $\olpi$ is 
defined to be $p^l-p^B$, is assumed to be due to concentrations of cations 
and anions and is assumed
to have a modified form of the van't Hoff equation (\ref{e1.17}),
$p^l-p^B = RT\phi(c^+ + c^-) + \pi_0$ where $\phi = 
\frac{\p(\ln a^w)}{\p(\ln x^w)}$, $a^w$ is the activity of the water
or solvent, $x^w$ is the molar concentration of the water, and $c^j$ are
the moles of ions per volume of fluid in porous material.   

Using $\olpi = p^l-p^B$ in equation (\ref{e3.70}) we see that 
$p^l-\olpi$ is the bulk phase pressure.    This form of the equation
can be derived from (\ref{e3.33}) if one uses (\ref{e3.40}) for the chemical
potential of the liquid phase, assumes the density of water is one,
and neglects the hydration of ions.

There is some question as to how well the ``osmotic pressure'', $p^l-p^B$,
which is physically the swelling pressure, can be approximated by the
van't Hoff equation which is used for species (and not swelling) 
osmotic pressure.

\section{Summary}

In this paper we show that the most general way to write
Darcy's law for swelling soils is in terms of gradients of chemical
potentials, see equation (\ref{e3.33}).  In this 
way one can float between using variables
such as pressure and moisture content and electro-chemical potentials
of either the vicinal fluid or a bulk fluid in electro-chemical
equilibrium.  Using this formulation tells us more easily the assumptions
used for other models, e.g.\  
Moyne and Murad \cite{murad06a} and Huyghe and Janssen \cite{huyghe97}. 

One clear consequence of this formulation is determining
when a pressure threshold gradient may exist.  This is the
pressure gradient that must be exceeded before flow is
observed.  A pressure gradient has been shown to exist
when whetting a previously dry sample, \cite{wei09,song10}, 
or for swelling soils such as clay \cite{sanchez07}. 
In this paper, we demonstrated 
that concentration gradients are negligible, if the pressure is 
the pressure of a reservoir  
in electro-chemical equilibrium with the swelling porous media, 
$p^B$, then there is
no pressure threshold gradient.  However if the pressure is
of the vicinal liquid within the porous media, $p^l$, then a threshold
gradient may exist - depending upon whether the swelling potential,
$\pi$ is nonzero.  See equations (\ref{e3.23}) and (\ref{e3.27b}) and
the discussion directly following them.    

This has implications in measuring the pressure - if one measures the
pressure within a swelling porous media with a device which takes
in (any) amount of fluid, then the fluid which is no longer
affected by the presence of the charged solid phase becomes
a bulk-phase fluid and is no longer at the same pressure as the
fluid within the swelling porous media.   Using such a device will not
indicate a critical pressure gradient.   One way that one can
obtain the pressure within a swelling porous material is by measuring
the overburden pressure - if the
solid phase supports no portion of the stress (i.e.\ it is
at the same pressure as the fluid) then the overburden pressure
is the pressure in the fluid and also in the solid - see for example,
Figures \ref{fig:f5a}, \ref{fig:f5b}, and \ref{fig:f5c}.

\bibliography{lynn,eqs}

{\bf Appendix A:  The change in chemical potential with respect 
to pressure}
\setcounter{equation}{0}
\def\theequation{A.\arabic{equation}}

In this appendix we go through the calculations to show that the partial
derivative of the chemical potential with respect to pressure while holding
concentrations and volume fraction fixed is constant if the specific densities
of each component is fixed.  We first show this is valid in terms of
extensive variables and then verify the result with our 
definition of chemical potential in terms of intensive variables.

In this section we suppress the notation for phase, $\a$, as the
definitions involved do not directly depend on which phase or
the volume fraction of the phase.    We assume there are $N$ constituents
making up the phase, and we define $C^j$ to be the mass fraction of
component $j$ with units (mass of j)(mass of phase).

\vspace{\baselineskip}

\noindent
{\bf Extensive Variables}

Let $G$ be the extensive Gibbs potential, $G=G(T,p,M^j,X)$ where
$p$ is pressure, $M^j$ is the mass of species $j$, and
$X$ is any other variable upon which the Gibbs potential depends, such
as the volume of the porous media.  We note that normally we write
$G$ as a function of the number of moles of species $j$, $N^j$, but
the ration of $M^j$ and $N^j$ is the molecular weight of $j$
(with units of mass of $j$ per mole of $j$), and since the molecular
weight is a constant this does not change the following results.  

The thermodynamic definition of chemical potential in units of
energy per unit mass is given by

\begin{eqnarray} \label{eA.1}
    \mu^j = \left.\frac{\p G}{\p M^j}\right|_{T,p,X}
\end{eqnarray}
Before deriving a Maxwell relation we use the total differential
to determine $\frac{\p G}{\p p}$:
\begin{eqnarray} \label{eA.2}
    dG &=& d(U-TS+pV)\nonumber \\ 
       &=& dU - TdS - SdT + pdV + Vdp \nonumber \\
       &=& \frac{\p U}{\p S}dS + \frac{\p U}{\p V} dV + \sum_{j=1}^N 
       \frac{\p U}{\p M^j}dM^j + \frac{\p U}{\p X}dX - TdS - SdT + pdV + Vdp 
         \nonumber \\
       &=& TdS - p dV + \sum_{j=1}^N \frac{\p U}{\p M^j}dM^j + 
        \frac{\p U}{\p X}dX - TdS - SdT + pdV + Vdp \nonumber \\ 
       &=&   \sum_{j=1}^N \frac{\p U}{\p M^j}dM^j + 
        \frac{\p U}{\p X}dX  - SdT  + Vdp  
\end{eqnarray}
where we used $T=\frac{\p U}{\p S}$, and $\frac{\p U}{\p V} = -p$ \cite{callen2}.

Now taking the partial of both sides with respect to $p$ keeping the appropriate
variables fixed we have:
\begin{eqnarray} \label{eA.3}
    \left.\frac{\p G}{\p p}\right|_{T,M^j,X} = V.
\end{eqnarray}

Now let's assume that the function $G$ is smooth enough so that mixed partials
commute.  Using (\ref{eA.3}) we have
\begin{eqnarray} \label{eA.4}
    &&  \frac{\p^2 G}{\p p \; \p M^j} = 
     \frac{\p^2 G}{\p M^j \; \p p}\nonumber \\
    && \frac{\p}{\p p} \left(\frac{\p G}{\p M^j}\right)= 
     \frac{\p}{\p M^j}\left(\frac{\p G}{\p p}\right) \nonumber \\ 
    &&   \left.\frac{\p \mu^j}{\p p}\right|_{T,M^k,X}
      = \left.\frac{\p V}{\p M^j}\right|_{T,p,M^k(k\neq j),X}.
\end{eqnarray}
Equation (\ref{eA.4}) tells us that the partial of the chemical potential with
respect to pressure is related to how the volume changes with the quantity of
$j$.

Recall that our goal is to show that if the intensive densities of the species
(so the mass of species $j$ per volume of species $j$)
are constant then so is $\frac{\p \mu^j}{\p p}$.   The units of $\rho^j$ are
mass of $j$ per volume of mixture.  So let's define the intensive density
to be $\rho_0^j$, which has units of mass of species $j$ per volume of
species $j$.  We use a subscipt 0 to emphasize the fact that it is not
equal to $\rho^j$ and that in what follows we consider $\rho_0^j$ to
be a constant.

Let $V= V^1+V^2+\dots+V^N$ be the volume of the mixture, where $V^j$ is the
volume of species $j$.  Then we have $M^j=\rho^j_0 V^j$.  With this,
(\ref{eA.4}) gives:
\begin{eqnarray} \label{eA.6}
     \left.\frac{\p \mu^j}{\p p}\right|_{T,M^k,X}
      &=& \left.\frac{\p V}{\p M^j}\right|_{T,p,M^l(l\neq j),X}
      = \sum_{k=1}^N \left.\frac{\p V^k}{\p M^j}\right|_{T,p,M^l(l\neq j),X}
      \nonumber \\
      &=& \sum_{k=1}^N \left.\frac{\p V^k}{\p (\rho_0^j V^j)}
        \right|_{T,p,V^l(l\neq j),X} \nonumber \\
        &=& \sum_{k=1}^N \frac1{\rho_0^j}\left.\frac{\p V^k}{\p V^j}
        \right|_{T,p,V^l(l\neq j),X} \nonumber \\
        &=& \frac1{\rho_0^j}.
\end{eqnarray}
So if the specific densities for every component is fixed then the dependence
of the chemical potential upon pressure is linear.  Note that if only
one species is incompressible, then we would need the additional assumption 
that the density of species $j$ is independent of the quantity (volume)
af all species (including $j$), in order for 
(\ref{eA.6}) to hold.

\vspace{\baselineskip}

\noindent
{\bf Intensive Variables}

We now go through the same argument in terms of intensive variables.  Since
the definition of chemical potential is relatively new, \cite{bmc2}, we go through
the calculations in detail to verify the same result holds.  The
chemical potential as defined in this paper in terms of the Helmholtz
potential, (\ref{e3.15}), is written with assumed independent variables,
$T,\ \rho^j,\ X$, where $X$ could be any other variable.  We would like
to determine the definition of chemical potential in terms of the
Gibbs potential, $g(T,p,C^j)$.  To do this we first look at the
definition of chemical potential in terms of $\tpsi = \tpsi(T,\rho,C^j,X)$.

\vspace{0.2in}

\noindent
{\bf Claim: } 
\begin{eqnarray} \label{eA.10}
p = \sum_{j=1}^N \rho \rho^j \left.\frac{\p \psi}{\p \rho^j} 
\right|_{T,\rho^l(l\neq j),X} = \rho^2 \left.\frac{\p \tpsi}{\p \rho}
\right|_{T,C^l(l=1,..,N-1), X}
\end{eqnarray} 

To show this result we begin with the equivalencies of the total differential
of the Helmholtz potentials:
\begin{eqnarray} \label{eA.11}
   && \tpsi(T,\rho,C^j,X) = \psi(T,\rho^k,X) 
   \nonumber \\
   &&d\tpsi(T,\rho,C^j,X) = d\psi(T,\rho^k,X) 
   \nonumber \\
   &&d\tpsi(T,\rho,C^j,X) = 
     \frac{\p \psi}{\p T}dT 
     + \sum_{k=1}^N\frac{\p \psi}{\p \rho^k} d\rho^k 
     + \frac{\p \psi}{\p X}dX .
     \nonumber
\end{eqnarray}

Now taking the partial with respect to $\rho$ on both sides keeping
the concentrations (and $T$ and $X$) fixed we have:
\begin{eqnarray} \label{eA.12}
    \left.\frac{\p \tpsi}{\p \rho} \right|_{T,C^j(j=1,\dots,N-1),X}
    &=& \sum_{k=1}^N \frac{\p \psi}{\p \rho^k} \left.\frac{\p \rho^k}{\p 
     \rho} \right|_{C^j}
      \nonumber \\
      &=& \sum_{k=1}^{N-1} \frac{\p \psi}{\p \rho^k}
           \left.\frac{\p(C^k\rho)}{\p \rho}\right|_{C^j}
        +  \frac{\p \psi}{\p \rho^N} \left.
        \frac{\p \left[(1-\sum_{l=1}^{N-1}C^l)\rho\right]}{\p \rho} 
          \right|_{C^j}
      \nonumber \\
      &=& \sum_{k=1}^{N-1} C^k \frac{\p \psi}{\p \rho^k}
        +  C^N \frac{\p \psi}{\p \rho^N} 
      \nonumber \\
      &=& \sum_{k=1}^{N} C^k \frac{\p \psi}{\p \rho^k}.
      \nonumber
\end{eqnarray}
Multiplying both sides by $\rho^2$ and using the fact that
$C^k \rho = \rho^k$ we get (\ref{eA.10}).

\vspace{0.2in}

\noindent
{\bf Claim: The chemical potential in terms of the Helmholtz
potential,
is given by} 
\begin{eqnarray} \label{eA.20}
    \mu^j &=& \left.\frac{\p(\rho\psi)}{\p \rho^j} \right|_{\rho^k (k\neq j)}
     = \psi + \frac{p}{\rho} - \sum_{k=1}^{N-1} C^k \left. \frac{\p \tpsi}{\p C^k} 
 \right|_{T,\rho,C^l(l\neq k), X} + \frac{\p \tpsi}{\p C^j}(1-\d^{jN}), 
 \nonumber \\ &&  \hspace{3.2in}
 \qquad j=1,\dots,N
\end{eqnarray} 
where $\d^{jN}$ is one if $j=N$ and zero otherwise.

We begin as we did in the previous claim by equating the two 
functions of Helmholtz potential, $\psi = \tpsi$ and looking at the
total differential.  We will then use the thermodynamic definition
of chemical potential given by (\ref{e3.15}),
$\mu^j = \left.\frac{\p (\rho\psi)}{\p \rho^j} \right|_{\rho^k (k\neq j)}$.
\begin{eqnarray} \label{eA.22}
    d\psi &= & d\tpsi 
    \nonumber \\
    &=& \left.\frac{\p \tpsi}{\p T}\right|_{\rho,C^k,X} dT
       +
     \left.\frac{\p \tpsi}{\p \rho}\right|_{T,C^k,X} d\rho
       +
     \sum_{k=1}^{N-1}\left.\frac{\p \tpsi}{\p C^k}
      \right|_{T,\rho,C^l(l\neq k),X} d C^k
      +
    \left.\frac{\p \tpsi}{\p X}\right|_{T,\rho,C^k} dX.
    \nonumber 
\end{eqnarray} 
Taking the partial derivative of both sides with respect to $\rho^j$ we have
\begin{eqnarray} \label{eA.24}
    \left.\frac{\p\psi}{\p \rho^j}\right|_{T,\rho^l (l\neq j),X}
    &=& \frac{\p \tpsi}{\p \rho}   
    \left.\frac{\p\rho}{\p \rho^j}\right|_{\rho^l(l\neq j)} 
    +
    \sum_{k=1}^{N-1}  \frac{\p \tpsi}{\p C^k}   
    \left.\frac{\p C^k}{\p \rho^j}\right|_{\rho^l(l\neq j)}.
\end{eqnarray} 
We now need to evaluate the terms $\p \rho / (\p \rho^j)$
and $\p C^k / (\p \rho^j)$:
\begin{eqnarray} \label{eA.26}
  \rho= \sum_{k=1}^N \rho^k \Rightarrow 
    \left.\frac{\p\rho}{\p \rho^j}\right|_{\rho^l(l\neq j)} = 1. 
\end{eqnarray} 
Also:
\begin{eqnarray} \label{eA.28}
 j=k: \qquad  \frac{\p C^j}{\p \rho^j} &=& \frac{\p}{\p \rho^j}
   \left(\frac{\rho^j}{\rho}\right) = \frac{\rho-\rho^j}{\rho^2}
   = \frac1{\rho} - \frac{C^j}{\rho}
   \nonumber \\
   j\neq k: \qquad  \frac{\p C^k}{\p \rho^j}
   &=& \frac{\p}{\p \rho^j} \left(\frac{\rho^k}{\rho}\right)
     = -\frac{\rho^k}{\rho^2} = -\frac{C^k}{\rho}
     \nonumber
\end{eqnarray} 
Substituting these results into (\ref{eA.24}) we get:
\begin{eqnarray} \label{eA.30}
    \left.\frac{\p\psi}{\p \rho^j}\right|_{T,\rho^l (l\neq j),X}
    &=& \frac{\p \tpsi}{\p \rho}   
    + \sum_{k=1}^{N-1} \frac{\p\tpsi}{\p C^k }
         \left(-\frac{C^k}{\rho}\right)
         + \frac{\p\tpsi}{\p C^j} \frac1{\rho}(1-\d^{jN})
         \nonumber
\end{eqnarray} 
So
\begin{eqnarray} \label{eA.32}
    \mu^j &=& \left.\frac{\p(\rho\psi)}{\p \rho^j} \right|_{\rho^k}
     = \psi + \rho 
    \left.\frac{\p\psi}{\p \rho^j}\right|_{T,\rho^l (l\neq j),X}
    \nonumber \\
    &=& \psi + \rho\left.\frac{\p \tpsi}{\p\rho}\right|_{T,C^l,X}
       - \sum_{k=1}^{N-1}C^k\left.\frac{\p \tpsi}{\p C^k}
       \right|_{T,\rho,C^l(l\neq k),X} +\frac{\p \tpsi}{\p C^j} (1-\d^{jN})
       \nonumber \\
    &=& \psi + \frac{p}{\rho}
       - \sum_{k=1}^{N-1}C^k\left.\frac{\p \tpsi}{\p C^k}
       \right|_{T,\rho,C^l(l\neq k),X} +\frac{\p \tpsi}{\p C^j} (1-\d^{jN}),
\end{eqnarray} 
where we used (\ref{eA.10}) and is the result of this claim.
In the above we note that $\psi=\tpsi$ as these represent the same 
quantities and we can choose the functional form of the Helmholtz potential.

Also note that we have
\begin{eqnarray} \label{eA.34}
    \mu^j-\mu^N = \left.\frac{\p\tpsi}{\p C^j}\right|_{T,\rho,C^l(l\neq k),X}
\end{eqnarray} 
which is the relationship derived using an exploitation of the
entropy inequality in \cite{bmc2}.

\vspace{0.2in}

\noindent
{\bf Claim: The chemical potential in terms of the Gibbs potential
is given by} 
\begin{eqnarray} \label{eA.40}
 \mu^j = g - \sum_{k=1}^{N-1} C^k \left. \frac{\p g}{\p C^k} 
 \right|_{T,\rho,C^l(l\neq k), X} + \frac{\p g}{\p C^j}(1-\d^{jN}), 
 \qquad j=1,\dots,N
\end{eqnarray} 
where $\d^{jN}$ is one if $j=N$ and zero otherwise.

We first derive the intensive equivalent to (\ref{eA.3}) by
beginning with the relationship between the Gibbs potential
and the Helmholtz potential,
\begin{eqnarray} \label{eA.42}
      g=\tpsi+\frac{p}{\rho},
      \nonumber
\end{eqnarray} 
where $g=g(T,p,C^k,X)$ for $k=1,\dots,N-1$.
Taking the total differential of both sides:
\begin{eqnarray} \label{eA.44}
    dg &=& d\tpsi
    +\frac1{\rho} dp - \frac{p}{\rho^2} d\rho
    \nonumber \\
    &=& \frac{\p \tpsi}{\p T} dT + \frac{\p\tpsi}{\p\rho} d\rho
      + \sum_{k=1}^{N-1} \frac{\p \tpsi}{\p C^k} dC^k
      + \frac{\p \tpsi}{\p X} dX
    +\frac1{\rho} dp - \frac{p}{\rho^2} d\rho
    \nonumber \\
    &=& \frac{\p \tpsi}{\p T} dT
      + \sum_{k=1}^{N-1} \frac{\p \tpsi}{\p C^k} dC^k
      + \frac{\p \tpsi}{\p X} dX
    +\frac1{\rho} dp 
\end{eqnarray} 
where we used (\ref{eA.10}) to cancel two terms in the last step.

To get the equivalent of (\ref{eA.3}) take the partial with respect
to $p$ on both sides and we have
\begin{eqnarray}\label{eA.46} 
    \left.\frac{\p g}{\p p} \right|_{T,C^j(j=1,\dots,N-1),X}
    = \frac1{\rho}
\end{eqnarray} 
which is consistent with (\ref{eA.3}) in the sense that if we
divide both sides of (\ref{eA.3}) by the total mass (and
the total mass is fixed) we get (\ref{eA.46}).   This remark just 
shows consistency.

To get the chemical potential in terms of the Gibbs potential begin
with (\ref{eA.44}),
\begin{eqnarray} \label{eA.48}
    \left.\frac{\p g}{\p C^j} \right|_{T,p,C^k(k\neq j),X}
       = \left.\frac{\p \tpsi}{\p C^j}\right|_{T,\rho,C^k(k\neq j),X},
\end{eqnarray}
and use the result from the previous claim, (\ref{eA.20}), to get
(\ref{eA.40}).

Two checks can be made on this result. If there is only
one component ($N=1$) then the chemical potential of the phase should
be the Gibbs potential, and the sum of the weighted chemical
potentials should be the Gibbs potential:
\begin{eqnarray}  \label{eA.50}
    \sum_{j=1}^N C^j\mu^j = g.
    \nonumber
\end{eqnarray} 
A few algebraic steps shows that both of these results hold.

Further, letting $j=N$ in (\ref{eA.40}), we get
\begin{eqnarray} \label{eA.52}
    \mu^N = g - \sum_{k=1}^{N-1} C^k \left.\frac{\p g}{\p C^k}
      \right|_{T,\rho,C^l(l\neq k),X}.
\end{eqnarray}

\vspace{0.2in}

\noindent
{\bf Claim: If each component of the phase is incompressible,
then the partial derivative of the chemical potentail with respect
to pressure is constant.
} 

We adapt the notation from the extensive results and let $\rho^j_0$
be the intrinsic density of component $j$ (mass of $j$ with
respect to volume of $j$).  For this claim we assume that
$\rho^j_0$ is constant for $j=1,\dots,N$.

We begin by showing a preliminary results using $C^j=\rho^j_0 v^j$
where $v^j$ is the volume of $j$ per unit mass
of the phase material, and is {\em not} $1/\rho^j$ which
has units of mass of $j$ per unit volume of phase material.  
Let $v = 1/\rho = \sum_{j=1}^N v^j$.  Then
\begin{eqnarray} \label{eA.60}
    \left. \frac{\p}{\p C^k}\left(\frac1{\rho}\right) 
    \right|_{C^l (l\neq k)} &=& \frac{\p}{\p C^k} (v)
    \nonumber \\
    &=& \frac{\p }{\p C^k} \left[\sum_{j=1}^N v^j \right]
    \nonumber \\
    &=& \frac{\p }{\p C^k} \left[\sum_{j=1}^N \frac{C^j}{\rho_0^j} \right]
    \nonumber \\
    &=& \frac{\p }{\p C^k} \left[\sum_{j=1}^{N-1} 
    \frac{C^j}{\rho_0^j} + \frac{(1-\sum_{j=1}^{N-1} C^j)}{\rho_0^N}
      \right]
    \nonumber \\
    &=& \frac1{\rho_0^k} - \frac1{\rho_0^N}
\end{eqnarray}

Now let's determine the partial derivative with respect to $\mu^N$ first.
Beginning with (\ref{eA.52}) and using (\ref{eA.46}) we have
\begin{eqnarray} \label{eA.62}
    \left.\frac{\p \mu^N}{\p p} \right|_{T,C^k,X}
    &=& \left.\frac{\p g}{\p p} \right|_{T,C^K,X}
      - \sum_{k=1}^{N-1} C^k \frac{\p^2 g}{\p p \p C^k}
      \nonumber \\
    &=& \frac{1}{\rho} 
    - \sum_{k=1}^{N-1} C^k \frac{\p}{\p C^k}\left(\frac{\p g}{\p p}\right)
      \nonumber \\
    &=& \frac{1}{\rho} 
    - \sum_{k=1}^{N-1} C^k \frac{\p}{\p C^k}\left(\frac{1}{\rho}\right)
      \nonumber \\
    &=& \frac{1}{\rho} 
    - \sum_{k=1}^{N-1} C^k \left[ \frac1{\rho^k_0} -\frac1{\rho^N_0}
      \right]
      \nonumber \\
      &=& \frac{1}{\rho}  + \frac{1-C^N}{\rho_0^N}
      - \sum_{k=1}^{N-1} \frac{C^k}{\rho_0^k} 
      \nonumber \\
      &=& \frac{1}{\rho}  + \frac1{\rho_0^N} - \frac{C^N}{\rho_0^N}
      - \sum_{k=1}^{N-1} v^k 
      \nonumber \\
      &=& \frac{1}{\rho}  + \frac1{\rho_0^N} - v^N
      - (v- v^N)  
      \nonumber \\
      &=& \frac{1}{\rho_0^N},
\end{eqnarray}
where we used result (\ref{eA.60}).

Now the rest is easy if we begin with (\ref{eA.34}) and (\ref{eA.48}):
$\mu^j = \mu^N + \p g/(\p C^j)$ for $j=1,\dots,N-1$:
\begin{eqnarray} \label{eA.64}
    \left.\frac{\p \mu^j}{\p p} \right|_{T,C^k,X}
     &=& \frac{\p}{\p p} \left(\mu^N + \frac{\p g}{\p C^j}\right)
     \nonumber \\
     &=& \frac{\p\mu^N}{\p p} + \frac{\p}{\p C^j}\frac{\p g}{\p p}
     \nonumber \\
     &=& \frac{1}{\rho_0^N} + \frac{\p}{\p C^j}\left(\frac1{\rho}\right)
     \nonumber \\
     &=& \frac{1}{\rho_0^N} + \frac1{\rho_0^k} - \frac1{\rho_0^N} 
     \nonumber \\
     &=& \frac1{\rho_0^k},
\end{eqnarray}
where we used (\ref{eA.46}) and (\ref{eA.62}) in going from line
2 to line 3. and (\ref{eA.60}) to go from line 3 to line 4.

And so we see that if the specific densities are constant (i.e.\ do not
change too much with the temperature and concentration fluctations of
the particular problem being considered), then the chemical potential
changes linearly with the {\em total} pressure.   This result
is generally used for liquids and not for gasses.

\bigskip 

{\bf Appendix B: Background Material on Chemical Potential }
\setcounter{equation}{0}
\def\theequation{A.\arabic{equation}}

This section contains material found in a standard textbook on physical
chemistry \cite{atkins,castellan}.  It is presented here for easy reference.

The chemical potential has three defining properties
(1) it is a scalar quantity representing the energy change as
the quantity of species is changed (partial derivative of a
potential with respect to quantity), (2) is a quantity which is
equal in two different phases at equilibrium, and (3) is the
generalized driving force for diffusion.

For a pure substance in a single phase, the chemical potential is equal 
to the Gibbs potential (per unit mole), $\overline{G}$.  
We first determine the Gibbs potential for
a single compenent, ideal gas that satisifies $pV = nRT$ where $V$ is the
volume, $p$ is the pressure, $n$ is the number of moles, $R$ is the 
gas constant,
and $T$ is the absolute temperature.  In this case, the Gibbs potential
is only a function of temperature and pressure: 
$G=G(T,p)$ and \cite{callen2,
atkins}
\begin{eqnarray*}
    d\overline{G} &=& \frac{\p \oG}{\p T} dT + \frac{\p \oG}{\p p} dp
    \\
    &=& -SdT + V dP.
\end{eqnarray*}
To determine how the Gibbs potential depends upon pressure,
integrate the above relationship from a reference state
$\oG_0(T_0,p_0)$ (where $p_0$ is the standard pressure) 
to a second state at constant temperature and
number of moles,
$\oG_0(T_0,p)$, and using the ideal gas relationship $p = RT/v_m$ (where
$v_m$ is the molar specific volume with units of volume per mole) we have
\begin{eqnarray*}
     \oG(T,p) &=& \oG_0(T,p_0) + \int_{p_0}^p v_m \; dP
     \nonumber \\
     &=& \oG_0(T,p_0) + \int_{p_0}^p \frac{RT}{p} \; dP
     \nonumber \\
     &=& \oG_0(T,p_0) + RT\ln\left(\frac{p}{p_0}\right).
\end{eqnarray*}
If we do not have an ideal gas then we replace the pressure by an effective
pressure, called the {\em fugacity}, $f$, and we have
\begin{eqnarray*}
     \oG(T,p) &=& \oG_0(T,p_0) + \int_{p_0}^p v_m \; dP
     \nonumber \\
     &=& \oG_0(T,p_0) + RT\ln\left(\frac{f}{p_0}\right),
\end{eqnarray*}
and in fact, this is the definition of fugacity.

For
a mixture of gases, define the partial pressure of species $j$ to
be $p^{g_j}=x^j p$ where $x^j$ is the molar fraction of species $j$ (moles
of $j$ per moles of mixture).  For a mixture of perfect gases (each
gas ideal and no interactions between species), the partial pressure
of species $j$ would actually be the pressure of species $j$
if no other species
were present (Dalton's law).  
Using the definition of partial pressure we have, for a component of 
a perfect mixture of gases:
\begin{eqnarray} \label{eB.13}
    \omu^{g_j}(T,p,x^j) = \omu^{g_j}_0(T,p_0) 
    + RT\ln\left(\frac{p^{g_j}}{p_0}\right),
\end{eqnarray}
where $\omu^{g_j}$ is the chemical potential of species $j$ in units of
energy per mole, and $p_0$ is the standard pressure (which is 1 if
pressure is measured in bars).
Using Dalton's law, 
we have
\begin{eqnarray} \label{eB.15}
    \omu^{g_j}(T,p,x^j) &=& \omu^{g_j}_0(T,p_0) + RT\ln\left(\frac{x^j p}{p_0}\right),
    \nonumber \\ \label{eB.16}
    &=& \omu^j_0(T,p_0) + RT\ln\left(\frac{p}{p_0}\right)
                        + RT\ln{x^j}.
\end{eqnarray}
Since $x_j$ is always between 0 and 1, we have that the last term is
always negative and so {\em the chemical potential of a component in 
a mixture is always less than the chemical potential of a pure substance}. 
We should note that the above result is only true for a mixture
of ideal gases.  This model breaks down if for example, there
are chemical reactions, the pressure is high, or there are strong
intermolecular forces between (or among) the different species.
Water vapor in the atmosphere is usually treated as an ideal gas,
with error in e.g.\ density calculations of less than 0.2\% 
(http://en.wikipedia.org/wiki/Density\_of\_air).  At high pressures
the perfect mixture assumption breaks down.

For a mixture of nonideal gases, the partial pressure must be 
replaced by the fugacity and we have
\begin{eqnarray*}
    \omu^{g_j}(T,p,x^j) = \omu^{g_j}_0(T,p_0) + RT\ln\left(\frac{f^j}{p_0}\right).
\end{eqnarray*}

For a liquid, the chemical potential is determined by using
the fact that the chemical potentials of one species in two phases
are equal at equilibrium.    Thus the chemical potential of a pure
liquid of an ideal component, $j$, (the component behaves as
an ideal gas in the gaseous state) is:
\begin{eqnarray} \label{eB.20}
    \omu^{l_j}_p = \omu^{g_j}_p = \omu^{g_j}_0 (T,p_0) + RT
    \ln\left(\frac{p^{g_j}_p}{p_0}\right),
\end{eqnarray}
where $\omu^{g_j}$ is the chemical potential of species $j$ in
the gaseous state in equilibrium with the liquid state, $\omu^{g_j}_0$
is the chemical potential of species $j$ in the gaseous state at
the same temperature but at standard pressure (1 bar), $p^{g_j}_p$
is the partial pressure of $j$ in equilibrium with pure $j$ in the
liquid phase, and $p_0$ is the standard pressure.
Now suppose we have a liquid mixture of ideal species.  Let $p^{g_j}$ be
the partial pressure of species $j$ in the gas phase.  Then
the chemical potential is:
\begin{eqnarray} \label{eB.21}
    \omu^{l_j} &=& \omu^{g_j}(T,p^{g_j}) = \omu^{g_j}_0 (T,p_0) + RT
    \ln\left(\frac{p^{g_j}}{p_0}\right)
     = \omu^{l_j}_p(T,p) + RT \ln\left(\frac{p^{g_j}}{p^{g_j}_p}\right),
     \nonumber \\
     &=& \omu^{l_j}_p(T,p) + RT \ln\left(a^j\right),
\end{eqnarray}
where we eliminated $\omu^{g_j}_0$ using (\ref{eB.20}) and
$a^j = p^{g_j}/p^{g_j}_p$ is the {\em activity} of component $j$.
We note that for water, the activity 
is the {\em relative humidity} divided by
100 (i.e.\ relative humidity not in percent form).
Since the partial pressure for a species in a mixture is usually
less than the partial pressure in the pure state (i.e.\ in the
case of water, the relative humidity is between 0 and 1), 
we see that the chemical
potential for a species in a mixture is generally lower than the 
chemical potential
of a pure species.  The above equation holds whether the
liquid solution is ideal or not.

Now if we have {\em an ideal} liquid solution, then Raoult's law is
satisfied (this is the definition of an ideal solution):
\begin{eqnarray*}
    p^{g_j} = x^j p^{g_j}_p
\end{eqnarray*}
where $p^{g_j}$ is the partial pressure of species $j$ in the
gas phase in equilibrium with the ideal solution, $x^j$ is the 
molar volume fraction in the liquid, and $p^{g_j}_p$ of the partial pressure 
of species $j$ in equilibrium with pure liquid $j$.  Using this relationship
we have
\begin{eqnarray*}
    \omu^{l_j} = \omu^{g_j}_0 (T,p_0) + RT
    \ln{x^j},
\end{eqnarray*}
where again, $x^j$ is the molar fraction of species $j$ in the
ideal solution.  An ideal solution is one in which the liquid molecules
in the mixture interact with all other species the same, i.e.\ there
is no difference between how species $i$ interacts with $i$ and how
species $i$ interacts with $j$.  This is a much stronger assumption
than assuming that the species behaves as an ideal gas in the gaseous phase.
Raoult's law is known to hold for a solvent when it is nearly pure.
For real solutions where species $j$ is at low concentration, instead
of $p^{g_j} = x^j p^{g_j}_p$ we replace $p^{g_j}_p$ with an empirical
constant (measured) so that $p^{g_j} = x^j K^j$ and this is referred
to as {\em Henry's law}.

Returning to (\ref{eB.21}) we want to use our knowledge from
Appendix A to express the chemical potential of a liquid in terms
of pressure.  Using the total differential and that
$\omu^\lj = m^j \mu^\lj$ we have:
\begin{eqnarray*}
    d\omu^\lj_p(T,p) &=& \frac{\p \omu^\lj_p}{\p p} dp +
     \frac{\p \omu^\lj_p}{\p T} dT,
     \nonumber \\ 
     &=& \frac{m^j}{\rho^\lj_0} dp +
     \frac{\p \omu^\lj_p}{\p T} dT,
\end{eqnarray*}
where no assumptions have been made.  Integrating both sides
from standard pressure to pressure, $p$, at a constant temperature we
have
\begin{eqnarray} \label{eB.25}
    \omu^\lj_p(T,p) - \omu^\lj_p(T,p_0) = \frac{m^j}{\rho^\lj_0} (p - p_0)
\end{eqnarray}
if $\rho^\lj$ is constant over the pressure range $p_0$ to $p$.
Combining this with (\ref{eB.21}) we have
\begin{eqnarray} \label{eB.27}
    \omu^\lj_p(T,p,x^j) = \omu^\lj_p(T,p_0) + \frac{m^j}{\rho^\lj_0} (p - p_0)
        + RT \ln(a^j).
\end{eqnarray}

So let us compare the difference between chemical potentials in the liquid
and gas phases, (\ref{eB.16}) and (\ref{eB.27}): 
\begin{eqnarray} \label{eB.30}
    \omu^{g_j}(T,p^g,x^{g_j}) 
    &=& \omu^j_0(T,p_0) + RT\ln\left(\frac{p^g}{p_0}\right)
    + RT\ln{x^{g_j}}.
    \\ \label{eB.31}
    \omu^\lj_p(T,p^l,x^\lj) &=& \omu^\lj_p(T,p_0) + 
    \frac{m^j}{\rho^\lj_0} (p^l - p_0) + RT \ln{a^j}.
\end{eqnarray}
Let's consider the case where $j$ is water, the gas phase is atmospheric
air, and water is the primary component of the liquid phase.  In this
case the ratio of pressures in the gas phase is close to 1 (atmospheric
pressure is close to 1 bar) and so the pressure term drops.  For
the liquid phase in which water is the primary component, the activity 
is approximately the molar concentration which is 1, and the term
involving the activity is negligible.  Thus we have
\begin{eqnarray} \label{eB.35}
    \omu^{g_j}(T,p^g,x^{g_j}) 
    &\approx& \omu^j_0(T,p_0) + RT\ln{x^{g_j}}.
    \\ \label{eB.36}
    \omu^\lj_p(T,p^l,x^\lj) &\approx& \omu^\lj_p(T,p_0) + 
    \frac{m^j}{\rho^\lj_0} (p^l - p_0)
\end{eqnarray}
that is, the chemical potential of water in the gas phase is primarily 
determined by the concentration, and the chemical potential of water
in the liquid phase is primarily determined by pressure.

If however there is something other than water in the liquid phase,
then the chemical potential of a component in the liquid phase
is determined by the pressure and the relative humidity in equilibrium
with the liquid phase.

%{\bf Rate of change of phase}   DRAFT

%Rate of evaporation is proportional to the difference between the
%chemical potentials.  From (\ref{eB.13}) and (\ref{eB.21}) we have
%\begin{eqnarray} \label{eB.40}
%    \omu^{g_j} &=& \omu^{g_j}_0(T,p_0) + 
%    RT \ln \left(\frac{p^{g_j}_{\textrm{current}}}{p_0}\right)
%      \\ \label{eB.42}
%      \omu^{l_j} &=& \omu^{g_j}(T,p^{g_j}_{\textrm{max\ for\ l}}) 
%        = \omu^{g_j}_0(T,p_0) 
%        + RT \ln \left(\frac{p^{g_j}_{\textrm{max\ for\ l}}}{p_0}\right)
%\end{eqnarray}
%So we have
%\begin{eqnarray} \label{eB.45}
%    \mu^{g_j} - \mu^\lj &=& RT \ln \left[ 
%    \frac{p_0}{p^{g_j}_{\textrm{max\ for\ l}}}
%    \frac{p^{g_j}_{\textrm{current}}}{p_0}
%    \right]
%    \\ \label{eB.48}
%    &=& RT \ln \left[ \frac{RH_{\textrm{air}}}{RH_l}  \right]
%\end{eqnarray} 
%where $RH_{\textrm{air}}$ is the relative humidity in the air 
%($P^{g_j}_{\textrm{curent}} / P^{g_j}_{\textrm{max}}$), and
%$RH_l$ is the relative humidity of liquid phase with solid
%($ P^{g_j}_{\textrm{max\ for\ l}}  / 
%P^{g_j}_{\textrm{max}} $).

\bigskip

{\bf Appendix C:\ \   Nomenclature}
\setcounter{equation}{0}
\def\theequation{C.\arabic{equation}}

In general, a superscript Greek letter indicates a macroscale quantity
from that phase.  Superscript minuscules indicate the constituent,
so that, e.g.\ $\bv_\a^j$ is the macroscopic velocity of constituent
$j$ in the $\a$-phase.  Subscript $b$ refers to the quantity in
the bulk or reservoir phase in electrochemical equilibrium with the 
vicinal fluid.
%A carrot over the symbol, $\widehat{\phantom{j}}$, is used to
%emphasize that the quantity represents a transfer
%from either another phase or from other constituents.

\begin{tabular}{ll}
    $a^\lj$ & activity of species $j$ defined to be the ratio of
    pressures $p^{g_j}/p_P^{g_j}$, [-] (\ref{e1.12}) \\
    $A$ & area [length$^2$] (\ref{e2.11}) \\
%$b^\aj$  & external entropy source for $j$th constituent in $\a$-phase, \\
%  & [entropy $\aj$/(mass $\aj$-time)]
%             \\
$C^\aj$  & mass concentration, $\rho^\aj/\rho^\a$, [-] (\ref{e1.10}) \\
%$\bd^\aj$ & symmetric part of $\bnab\bv^\aj$ [1/time] \\
%$e^\aj$ & internal energy, [energy $\aj$/mass $\aj$]  \\
%$\he^\aj_\b$ & rate of mass exchange from $\b$-phase to $\a$-phase of
%          $j$th\\ & constituent, [1/time] \\
$\bE$ & electric field [force/charge] (\ref{e3.11})
\\
$\bEs^s$ & macroscale strain tensor of solid phase, [-] (\ref{e3.11}) 
\\
$\bF$ & deformation tensor [-] (\ref{e3.70}) \\
%$\bg$ & gravity, [length/time$^2$] \\
$G^\a$ & Gibbs free energy, [energy of $\a$-phase/mass $\a$] 
   (\ref{e2.11}), (\ref{e3.25})\\
%$h^\aj$ & external supply of energy, [energy of $\aj$/(mass $\aj$-time)]  \\
%$\hbi^\aj$ & gain of momentum of constituent $j$ of phase $\a$ due
%     to mechanical\\ & interactions 
%      with other species within the same phase, [length/time$^2$]
%     \\
$\bj^j$ & ion flux [length/time] Section \ref{s5.1} \\
$m^j$ & molar mass [mass / mole of $j$] (\ref{e1.10}) \\
%$\bn_\a$ & unit outward normal to phase $\a$ \\
$n^j$ &volumetric ion concentration [volume ion $j$/volume of solvent]
   Section \ref{s5.1} \\
$N_i$ & number of moles of species $i$ [moles] (\ref{e2.11}) \\
$p$ & pressure [force/area] (\ref{e1.10}) \\
$p^\a$ & classical pressure (1/3 trace of cauchy stress tensor at rest), 
[force/length$^2$]  (\ref{e3.12})\\
  $p^{g_j}$ & partial pressure of species $j$ in the gas 
     phase (\ref{e1.10}) \\
  $p^{g_j}_P$ & maximum partial pressure of species $j$ in the gas phase 
    (\ref{e1.10})\\
%$\olp^\a$ & thermodynamic pressure, [force/length$^2$] \\
$q_e^\a$ & charge density of phase $\a$, $\sum_{j=1}^N
  \rho^\aj z^\aj$ [charge $\a$/volume $\a$]
   (\ref{e3.20}) \\
%$\bq^\aj$ & heat flux of constituent $j$ in $\a$-phase, 
%   [energy $\aj$/(length$^2$-time)]\\
%$\hQ^\aj_\b$ & gain of energy of constituent $j$ in phase $\a$ due
%  to non-chemical, \\ & non-mechanical transfer
%      with the other phase, \\ & [$\aj$ energy / (mass $\aj$-time)] \\
%$\hQ^\aj$ & energy gained by constituent $j$ in phase $\a$ due
%    to non-chemical, \\
%    & non-mechanical 
%    interactions with other constituents \\
%    & within phase $\a$
%       [energy of $\aj$/(mass $\aj$-time)]  \\
$\bQ^\lj$ &coefficient for generalized Fick's law (\ref{e3.22}) \\
$r^\lj$ & coefficient for capturing ion hydration effects (\ref{e3.20}) \\
%$\hr^\aj$ & rate of $j$th constituent mass gained within phase $\a$,
%  [1/time] \\
    $R$ & Universal gas constant, [Force-length/(degree-moles)] 
      (\ref{e1.10}) \\
%$\bR^l$ & resistivity tensor, a linearization
%  coefficient, see (\ref{e3.63}) \\
%$\bt^\aj$ & stress tensor of $j$th constituent in $\a$-phase,
%   [force/length$^2$] \\
%$\bt^\a$ & stress tensor of $\a$-phase =
%     $\sum_{j=1}^N [\bt^j_\a - \rho_\a^j\bu^j_\a \bu^j_\a]$\\
%\end{tabular}{ll}
%\begin{tabular}{ll}
$t$  & time \\
$T$  & absolute temperature [degree] (\ref{e1.10})\\
%$\hbT^\aj_\b$ & gain of momentum of phase $\a$ due to mechanical interactions
%    with \\ & the other phase
%   [length/time$^2$] \\
%$\bu^\aj$ & diffusive velocity, $\bv_\a^j-\bv_\a$ \\
%$v^\a$ & specific volume, [volume $\a$/mass $\a$] \\
$\bv^\aj$ & velocity  of $j$th constituent in phase $\a$, (\ref{e3.21})
      [length/time] \\
$\bv^\a$  & mass-averaged velocity of phase $\a$,
$\sum_{j=1}^N C^\aj\bv^\aj$, [length/time] (\ref{e3.11}) \\
$\bv^{\aj,\a} $ & diffusive velocity, $\bv^\aj-\bv^\a$ [length/time] 
  (\ref{e3.11}) \\
$\bv^{l,s}$ & velocity of liquid relative to solid phase, $\bv^l-\bv^s$,
   [length/time] (\ref{e3.11}) \\
   $x^\lj$ & molar fraction of j$^{th}$ species in phase $l$, [-] (\ref{e1.16})
\\
$z^\aj$ & fixed charge density associated with $\aj$, [ charge $\aj$/mass
$\aj$] (\ref{e3.11}), (\ref{e3.15b}) \\
%$V^\a$ & total volume of phase $\a$, [volume $\a$] \\
%$V$ & total volume $=V^s+V^l$, [volume] \\
%$\d V$ & Representative Elementary Volume (REV)\\
%$\d V_\a$ & portion of $\a$-phase within REV \\
%$\bw_\ab^j$ & velocity of $j$th constituent in the interface 
%         [length/time]\\
%$\bx$  & Eulerian coordinates \\
%$\gamma_\a$ & indicator function for phase $\a$ \\
$\e^\a$ & volume fraction of $\a$-phase in Representative Elementary
\\ &  Volume (REV), $|\d V_\a|/|\d V|$, [-] (\ref{e3.11}) \\
$\tilde{\e}$ & dielectric constant of solvent [-] Section \ref{s5.1} \\
$\tilde{\e}_0$ & vacuum permittivity [charge$^2$/Force-Length$^2$] 
  Section \ref{s5.1} \\
%$\eta^\aj$ & entropy density, [$\aj$ entropy/(mass $\aj$-time)]  \\
%$\heta^\aj$ & entropy gain of $j$th constituent in $\a$-phase
%        due to non-mass transfer\\ & interactions 
%      with other constituents within
%        phase $\a$ \\ &
%       [$\aj$ entropy/(mass $\aj$-time)]\\
%$\lam^\aj$ & Lagrange multiplier for continuity equation of $j$th
%   constituent \\ & in phase $\a$ \\
%$\lam^\a$ & thickness of layer $\a$ in the
%   swelling porous medium with a \\ & non-interacting
%   solid phase. \\
%$\hLam^\aj$ & entropy production density, [$\aj$ entropy/(mass $\aj$-time)] \\
\end{tabular}

\begin{tabular}{ll}
$\lambda^s$ & thickness of montmorillonite clay mineral, [length]
   (\ref{e1.18})
  \\
$\lambda^l$ & thickness of vicinal liquid of montmorrilonite clay, [length]
   (\ref{e1.18}) \\
$\mu^\aj$  & chemical potential of $j$th constituent in phase
        $\a$ \\
        & [energy $\a$/ mass $\aj$] (\ref{e1.10}), (\ref{e3.15})\\
$\tmu^\aj$  & electro-chemical potential of $j$th constituent in phase
        $\a$ \\
        & [energy $\a$/ mass $\aj$] (\ref{e3.15b})\\
$\omu^\aj$  & chemical potential of $j$th constituent in phase
        $\a$ \\
        & [energy $\a$/ mole $\aj$] in Appendix B (\ref{e3.15b})\\
        & [energy $\a$/ molecule $\aj$] in Section \ref{s5.1} \\
$\pi^\lj$ & osmotic pressure, [force/area] (\ref{e1.12}) \\
$\pi^\a$ & swelling pressure, [force/area] (\ref{e3.14}) \\
$\phi$ & electric field potential, $\bE = -\bnab \phi$,
    [force/charge-length] (\ref{e3.15b}) \\
%$\Psi^{\aj}$ & Helmholtz free energy density of $j$th constituent in
%     $\a$-phase \\ &  $e^\aj-T\eta^\aj$, 
$\Psi^{\a}$ & Helmholtz free energy density of $\a$-phase,
     \\ & $e^\a -T\eta^\a$
   [energy $\a$ / mass $\a$] (\ref{e3.12})\\
$\rho^\aj$ & density of $j^{\rm th}$ constituent in phase $\a$,
    $C^\aj\rho^\a$,
    [mass $\aj$/ volume $\a$]  (\ref{e3.11}) \\
$\orho^\aj$ & specific mass density of  phase $\a$, 
[mass $\aj$/ volume $\aj$] (\ref{e1.13b}) \\
$\rho^\a$ & averaged mass density of  phase $\a$, $\sum_{j=1}^N \rho^\aj$,
   [mass $\a$/ volume $\a$] \\
%$\bphi^\aj$ & entropy flux of $j^{\rm th}$ constituent of phase $\a$, 
   %[entropy $\aj$/ (length$^2$-time)]\\
%$\hPhi^\aj_\b$ & entropy gained by $j$th constituent in $\a$-phase due
%       to non-mass \\ & transfer  interactions 
%     with the other phase [entropy /(mass $\aj$-time)] \\

\end{tabular}

\end{document}